\documentclass[lettersize,journal]{IEEEtran}
\ifCLASSINFOpdf

\else

\fi

\usepackage{CJK}
\usepackage{pifont}

\usepackage{algorithmic}
\usepackage{amsmath}
\usepackage{amssymb}
\usepackage{psfrag}
\usepackage{graphicx}
\usepackage{epstopdf}
\usepackage{multirow}
\usepackage{stfloats}
\usepackage{cite}
\usepackage{subcaption}
\usepackage{color}
\usepackage[normalem]{ulem}
\usepackage{arydshln}
\usepackage{enumerate}
\usepackage{bbm}\usepackage{verbatim}

\usepackage{url}
\newtheorem{theorem}{\bf{Theorem}}[section]

\newcommand{\bm}[1]{\mbox{\boldmath{$#1$}}}

\usepackage[linesnumbered,ruled,vlined]{algorithm2e}

\begin{document}
\title{Radiation Footprint Control in Cell-Free Cooperative ISAC: Optimal Joint BS Activation and Beamforming Coordination}

\author{Jie Chen, \IEEEmembership{Member, IEEE}, and Xianbin Wang, \IEEEmembership{Fellow, IEEE}
\thanks{
Manuscript submitted to IEEE Transactions on Communications on 15 January 2025; revised on 14 April 2025, and 29 May 2025; accepted on 8 July 2025.
This work was supported in part by the Natural Sciences and Engineering Research Council of Canada (NSERC) Discovery Program under Grant RGPIN-2024-05720 and in part by the Canada Research Chair Program under Grant CRC-2023-00336.  (Corresponding author: Xianbin Wang.)

J. Chen and X. Wang are with the  Department of Electrical and Computer Engineering, Western University, London, ON N6A 5B9, Canada (e-mails: jiechen@ieee.org, xianbin.wang@uwo.ca).
}

}
 \maketitle
\begin{abstract}
Coordinated beamforming across distributed base stations (BSs) in cell-free wireless infrastructure can efficiently support integrated sensing and communication (ISAC) users by enhancing resource sharing and suppressing interference in the spatial domain.
However, intensive coordination among distributed BSs within the ISAC-enabled network
poses risks of generating substantial interference to other coexisting networks sharing the same spectrum, while also incurring elevated costs from energy consumption and signaling exchange.
To address these challenges, this paper develops an interference-suppressed and cost-efficient cell-free ISAC network, which opportunistically and cooperatively orchestrates distributed radio resources to accommodate the competing demands of sensing and communication (S\&C) services.
Specifically, we conceive a radiation footprint control mechanism that autonomously suppresses interference across the entire signal propagation space to safeguard other networks without exchanging channel knowledge signaling. Then, we propose joint BS activation and beamforming coordination to dynamically activate appropriate BSs and orchestrate their spatial beams for service provisioning.
Building upon this framework, we formulate a cost-efficient utility maximization problem that considers individual S\&C demands and location-dependent radiation footprint constraints. Since this results in a non-convex optimization problem, we develop a monotonic optimization embedded branch-and-bound (MO-BRB) algorithm to find the optimal solution. Additionally, we apply a low-complexity iterative method to obtain near-optimal solutions.
Finally, simulation results validate the effectiveness of the proposed algorithms.
\end{abstract}


\IEEEpeerreviewmaketitle

\section{Introduction}

With the rapid convergence of wireless communications and vertical applications, supporting capabilities beyond communications, particularly sensing and positioning, is considered an essential aspect of next-generation wireless networks.
Consequently, integrated sensing and communication (ISAC) technology is expected to play a pivotal role in enabling diverse use cases identified in the sixth-generation (6G) vision outlined in the International Mobile Telecommunications (IMT)-2030 \cite{series2023imt}.
Specifically, ISAC can unify both sensing and communication (S\&C) functions into an integrated platform with shared signal waveforms, energy, and spectrum resources, thus efficiently fulfilling the stringent  S\&C demands concurrently \cite{wei2023integrated,zhang2021enabling,liu2020joint,dong2022sensing,hu2024digital}.
Owing to its significant benefits and promising prospects, ISAC has been extensively investigated in recent years by both the industrial and academic communities.


One fundamental challenge of ISAC-enabled systems is how to allocate the constrained radio resources effectively in supporting competing S\&C needs.
Consequently, many existing studies have focused on investigating the beamforming schemes in single base station (BS)-based ISAC systems to achieve various trade-offs between S\&C performances.
Specifically, in cases considering a single transmission frame \cite{hua2023mimo,ni2021multi,hua2023optimal,wen2023efficient,he2023full,johnston2022mimo,chen2022generalized}, the beamforming matrices were optimized to achieve trade-offs between the communication rate and various sensing metrics, i.e.,  Cram\'{e}r-Rao bound (CRB)  in \cite{hua2023mimo}, sensing mutual-information rate in \cite{ni2021multi}, or beam pattern matching error in \cite{hua2023optimal}.
Also, the signal-to-interference-plus-noise ratios for S\&C signals were balanced during the beamforming design by considering per-antenna power constraint in \cite{wen2023efficient} and total power constraint in~\cite{he2023full,johnston2022mimo,chen2022generalized}.
For scenarios involving successive or multiple transmission frames \cite{liu2020radar, yuan2020bayesian, liu2022learning, chen2023impact, chen2024learning}, beamforming matrices and power allocation were optimized to balance communication rate and sensing CRBs.
These optimizations leveraged predicted kinematic parameters of targets in the future transmission frames by applying extended Kalman filtering \cite{liu2020radar}, Bayesian inference \cite{yuan2020bayesian}, and deep learning networks \cite{liu2022learning}.
Furthermore, channel temporal correlation was leveraged in \cite{chen2023impact,chen2024learning} to reduce the channel estimation overhead during the trade-off design.
However, these ISAC beamforming schemes relying on a single BS exhibit inherent limitations in coverage, user capacity, transmission reliability, and resource efficiency due to the constraints in power budget and spatial diversity, as well as the challenges in adapting to the rapidly changing wireless communication environment.

Recently, the cell-free cooperative distributed ISAC network has emerged as a promising paradigm for addressing the limitations of single-BS ISAC networks.
This is because the cell-free cooperative mechanism enables the joint exploitation of distributed BS antenna arrays to enhance spatial diversity, while leveraging transceiver geographical relationships to improve spatial-domain resource sharing for heterogeneous service provisioning.
Although cooperative distributed radar systems have been studied over the past few decades \cite{li2022joint}, there are very few studies focusing on cooperative distributed ISAC networks due to distinct operational objectives, environments, and resource constraints.
Specifically,  the  S\&C performance trade-off is achieved by optimizing power allocation in \cite{ahmed2019distributed} and spectrum sharing in \cite{wu2022resource} for a distributed ISAC network with single-antenna BSs.
As for multiple-antenna BS cases, the joint optimization of power allocation and access point operation modes for S\&C tasks was proposed in \cite{elfiatoure2023cell} to maximize the communication spectral efficiency with sensing mainlobe-to-average-sidelobe ratio constraints.
Then, the joint active and passive beamforming was investigated in a bi-static ISAC system to maximize sensing performance while meeting communication demands \cite{kong2024signal}.
Moreover, \cite{mao2023beamforming} extended these solutions to imperfect channels and optimized beamforming to balance beam pattern mismatch and ergodic rate.

However, deploying dense BSs incurs new challenges in cell-free cooperative ISAC networks.
Firstly, coordinating dense distributed BSs substantially raises operational costs related to power consumption, encoding/decoding complexity, and signaling overhead for BS management or information exchanges among the BSs.
Hence, one main challenge of distributed systems is how to dynamically activate the proper BSs for cost-efficient service provisioning.
Specifically, the joint BS activation and coordinated beamforming design in conventional distributed multi-user multiple-input single-output (MISO) systems was studied in \cite{bi2019joint} to minimize BS power consumption for downlink cellular networks, and the Benders’ decomposition was applied to find its optimal solution.
Then, the joint BS activation and user association design was studied in \cite{zhou2023joint} to maximize the long-term transmission rate for ultra-dense heterogeneous networks.
However, these studies were designed for conventional communication networks without considering the disparities between S\&C service regions and channel characteristics, thus causing significant performance degradation when applied to cooperative ISAC networks.

Moreover, fully centralized control in large-scale cell-free deployments is impractical due to high complexity, signaling overhead, and limited scalability. Consequently, such systems are partitioned into multiple independent coexisting homogeneous or heterogeneous networks, and the spectrum is further shared among them to enhance spectral efficiency. However, it incurs a new challenge:  most existing coordination schemes address only intra-network interference and are ineffective at suppressing cross-network interference in the absence of cross-network interference channel knowledge.
To tackle this challenge, we propose to enable individual networks in the shared spectrum to autonomously limit their radiation footprint within their entire service areas. This strategy ensures dependable and resilient operation for each network without the need for extensive channel information sharing among diverse networks or causing significant interference to one another. This mechanism for suppressing potential interference is called radiation footprint control.
In our previous work \cite{wang2008radiation}, we established a transmission power negotiation signaling between the transmitter and receiver, thus achieving the minimum transmission power with the guaranteed QoS, consequently minimizing interference among nearby users.
However, that design targeted conventional point-to-point communication systems and is not applicable to cell-free ISAC networks.

To address the above challenges in the cell-free cooperative distributed ISAC network, this paper proposes an autonomous and opportunistic distributed resource orchestration framework to accommodate competing S\&C demands.
First, to reduce the coordination cost, a joint BS activation and beamforming coordination scheme is developed for cost-efficient service provisioning.
Second, to further suppress cross-network interference without requiring cross-network channel information exchange, a radiation footprint control mechanism is introduced to regulate each network's interference impact within the service area under shared spectrum.
To highlight the main contributions, we summarize the paper as follows:
\begin{itemize}
\item
We propose an interference-suppressed and cost-efficient cell-free distributed ISAC network by opportunistically leveraging distributed BS power and spatial resources to address multiple concurrent individual  S\&C demands.
Specifically, we first conceive a novel radiation footprint control mechanism to autonomously suppress interference throughout the signal propagation space for safeguarding other networks sharing the spectrum without exchanging interference channel knowledge.
Besides, we propose a joint BS activation and beamforming coordination scheme to dynamically schedule partial BSs for activation and orchestrate their power and spatial beamforming resources for high-quality S\&C service provisioning and operational cost reduction.

\item
Mathematically, we characterize the proportional cost-efficient  S\&C utility by deriving the transmission rate for communication service and the CRB for sensing service. Then, we maximize this utility by jointly optimizing the BS activation and spatial beamforming while considering the individual S\&C service demands, the location-varied radiation footprint interference constraints, and other practical constraints.
However, this problem falls into the mixed-integer nonlinear programming (MINLP) family, which is challenging to solve.

\item
We develop both optimal and sub-optimal algorithms to address the formulated problem efficiently.
Specifically, we convert the original problem into an equivalent but tractable formulation and investigate its monotonicity properties to facilitate algorithm development.
Following this, we propose an efficient monotonic optimization-embedded branch-and-bound algorithm (MO-BRB) to find its optimal solution.
Moreover,  we leverage the successive approximation (SCA) algorithm to find near-optimal solutions within polynomial time complexity.
\end{itemize}

Organizations: Section II introduces the system model, and Section III analyzes the associated system performance metrics.
Section IV formulates the joint BS activation and beamforming coordination problem.
Then, Section V and Section VI develop the optimal and low-complex near-optimal algorithms, respectively.
Finally, Section VII provides the simulation results, and Section VIII concludes the paper.

Notations: Scalars, vectors, and matrices are denoted by lowercase ($a$), bold lowercase ($\mathbf{a}$), and bold uppercase ($\mathbf{A}$), respectively. The transpose, conjugate transpose, trace, Frobenius norm, and rank are denoted by $(\cdot)^{\rm T}$, $(\cdot)^{\rm H}$, ${\rm Tr}(\cdot)$, $|\cdot|$, and ${\rm rank}(\cdot)$, respectively.
The $i$-th and $(i,j)$-th elements of a vector and matrix are denoted by $\left[\bf a\right]_i$ and $\left[\bf A\right]_{i,j}$, respectively.
 Besides, ${\bf I}_M$ denotes the {\mbox{$M$-by-$M$}} identity matrix; ${\mathbb E}(\cdot)$  is the expectation operator; $\mathrm{diag}(\cdot)$  constructs a square diagonal matrix from a vector; $\Re\left\{\cdot\right\} $ denotes the real part; and  ${\rm{rect}}\left(x \right)$ is the rectangular function, i.e., ${\rm{rect}}\left(x \right)=1$ if $0\le x\le T$, otherwise ${\rm{rect}}\left(x \right)=0$. Finally, $\left\lfloor \cdot \right\rfloor$ and $\left\lceil \cdot \right\rceil$ denote the flooring and ceiling functions, respectively.

\begin{figure}[t]
\center
\includegraphics[width=0.46\textwidth]{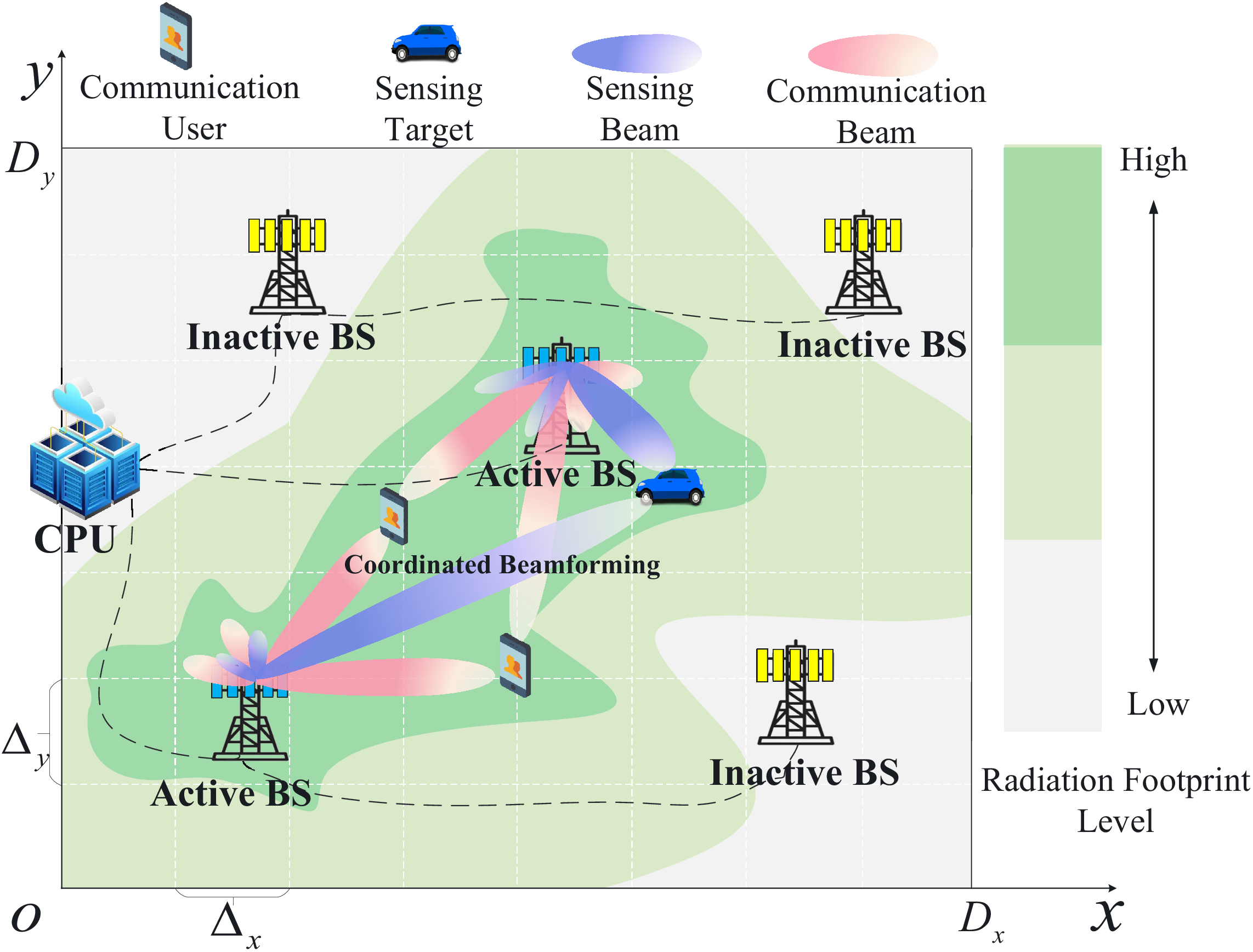}
\caption{A cell-free cooperative distributed MISO ISAC network where only partial BSs are opportunistically scheduled to be active and jointly provide S\&C services through coordinated ISAC beams. } \label{fig1}
\end{figure}
\section{System Model}

As shown in Fig. \ref{fig1}, we consider a cooperative distributed cell-free MISO ISAC network in a rectangular area bounded by $(0, 0)$ and $(D_x, D_y)$. This system comprises one central processing unit (CPU), $B$ full-duplex roadside BSs,  $K$ communication users, and  $Q$ radar point targets.
Each BS is equipped with $N_{\rm tx}$ antennas, while each user is equipped with a single antenna.
To reduce the operational costs of BSs, the CPU dynamically activates a subset of BSs for S\&C service provisioning, while deactivating others for cost savings.
Here, the BSs collect the CSI of all communication users through uplink channel estimation and relay it to the CPU.
Assuming perfect CSI, the CPU jointly performs BS activation and beamforming coordination for efficient service provisioning.
Specifically, each scheduled active BS cooperatively transmits ISAC signals to facilitate individual information transmission to communication users, while simultaneously relaying received echo ISAC signals---generated by itself or other BSs---to the CPU via optical links for joint target sensing.

\subsection{ISAC Signal Model}
Let the binary variable  ${ a}_b\in\left\{0,1\right\}$ denote whether BS $b$ is active, i.e., ${a}_b = 1$ indicates that BS $b$ is scheduled to be active, and ${a}_b = 0$ otherwise.
Then, motivated by the inherent resistance to multipath fading and flexibility in the system design of orthogonal frequency-division multiplexing (OFDM)  signals \cite{chen2024learning}, we adopt OFDM-based ISAC signals for S\&C service provisioning.
Besides, we design the ISAC signals as the sum of independent S\&C signals with dedicated beamforming vectors/matrices to facilitate flexible design and enhance the ability to meet distinct S\&C demands~\cite{ding2022joint,liu2020joint1}.
Let $L$ and $M$ denote the number of OFDM symbols and subcarriers, respectively. Then, the modulated $l$-th symbol on the $m$-th subcarrier at BS $b$ is given by
 \begin{align}
 {\bf{\hat s}}_{b}^{ml} =a_b\left( \sum\nolimits_{k = 1}^K {{\bf{w}}_{bk}^{\rm{C}}\hat {  s}_{kml}^{\rm{C}} + {\bf{W}}_{b}^{\rm{R}}{\bf{\hat s}}_{bml}^{\rm{R}}}\right)\in{\mathbb C}^{N_{\rm tx}\times1},\label{eq2}
 \end{align}
where the superscripts ``C'' and ``R'' indicate that the terms are related to communication and radar sensing modules, respectively.
Also, ${\bf{w}}_{bk}^{\rm{C}} \in {\mathbb C}^{{N_{{\rm{tx}}}} \times1}$ and ${\bf{W}}_{b}^{\rm{R}}{ \in {{\mathbb C}^{{N_{{\rm{tx}}}} \times N_{\rm tx}}} }$ are the corresponding dedicated beamforming vectors/matrices of the communication signal  $\hat { s}_{kml}^{\rm{C}}$ and the sensing signal $\hat {\bf s}_{bml}^{\rm{R}}$ by BS $b$, respectively.
Besides, we assume that $\hat s_{kml}^{\rm{C}}$ and $\hat  {\bf s}_{bml}^{\rm{R}} \in {\mathbb C}^{{N_{{\rm{tx}}}} \times1}$ are independent complex Gaussian variables with zero mean and covariances $1$ and $\mathbf{I}_{N_{\rm tx}}$, respectively.

Then, the OFDM-based ISAC signals generated by BS $b$ for information transmission and target sensing are given by
\begin{align}
{{\bf{s}}_b}\left( t \right) = \sum\limits_{m = 0}^{M - 1} {\sum\limits_{l = 0}^{L - 1} {{\bf{\hat s}}_b^{ml}{e^{{\rm{j}}2\pi m{\Delta _f}\left( {t - {T_{\rm cp}} - lT} \right)}}{\rm{rect}}\left( {t - lT} \right)} }
\label{eqisacsignal},
\end{align}
where ${\Delta_f} = \frac{1}{T_o}$ is the subcarrier bandwidth; $T_o$ and $T_{\rm cp}$ denote the OFDM symbol duration and cyclic prefix duration, respectively; and $T = T_o + T_{\rm cp}$ is the total symbol duration including the cyclic prefix.

\subsection{ Channel Model}
In this subsection, we develop mathematical models of the propagation channel for radiation footprint evaluation, the communication channel for communication rate characterization, and the target mobility for sensing accuracy analysis.

\subsubsection{Propagation Channel}
To characterize the radiation footprint of all BSs across the service area, it is necessary to model the propagation channel from each BS within the entire service provisioning area. Therefore,  we partition the entire service area into $X \times Y$ rectangle sub-regions with side lengths of $\Delta_x$ and $\Delta_y$ meters, where $X=\left\lceil {\frac{{{D_x}}}{{{\Delta _x}}}} \right\rceil $ and $Y=\left\lceil {\frac{{{D_y}}}{{{\Delta _y}}}} \right\rceil $ represent the number of sub-regions along the x-axis and y-axis, respectively.  Then, we denote the index set of all sub-regions by ${\cal S}=\left\{ {\left( {{x},{y}} \right)\left| {1 \le {x} \le {X},1 \le {y} \le {Y}} \right.} \right\}$.
Upon denoting  the positions of  BS $b$   and the central point of  sub-region $(x,y)$ by ${\bf{d}}_{b}^{\rm{B}} = {\left[ {x_{b}^{\rm{B}},y_{^b}^{\rm{B}}} \right]^{\rm T}}$ and ${{\bf{d}}_{xy}} = {\left[ {\left( {x - 0.5} \right){\Delta _x},\left( {y - 0.5} \right){\Delta _y}} \right]^{\rm{T}}}$, respectively, the propagation channel from   BS $b$ to  sub-region $(x,y)$ on each subcarrier is  as a Rician fading channel, i.e.,
\begin{align}
 {{\bf{g}}_{bxy}} = {\bf{g}}_{bxy}^{\rm{L}} + {\bf{g}}_{bxy}^{{\rm{NL}}} \in {{\mathbb C}^{{N_{{\rm{tx}}}} \times 1}},\label{eqaa3}
\end{align}
where  ${\bf{g}}_{bxy}^{\rm{L}}$ and ${\bf{g}}_{bxy}^{{\rm{NL}}}$  are the Line-of-Sight (LoS) and Non-LoS (NLoS) channel components, respectively.
Letting the Rician factor be $\kappa$,   ${\bf{g}}_{bxy}^{\rm{NL}}$ can be modeled as an independent complex Gaussian vector with   zero mean and covariance
$\bar\kappa _{bxy}{\bf I}_{N_{\rm tx}}$, where $\bar\kappa_{bxy} = \frac{1}{\kappa + 1} \left\| {\bf d}_b^{\rm B} - {\bf d}_{xy} \right\|^{-\xi}$ and $\xi$ is the path loss exponent.
Then, assuming the half-wavelength antenna spacing, the LoS component is  ${\bf{g}}_{bxy}^{\rm{L}} = {\bf{v}}\left( {{\bf{d}}_b^{\rm B},{{\bf{d}}_{xy}}} \right)$, where 
\begin{align}
{\bf{v}}\left( {{\bf{d}}_b^{\rm{B}},{{\bf{d}}_{xy}}} \right)
=  \frac{{\sqrt \kappa  {{\left[ {1, \ldots ,{e^{ - {\rm{j}}\pi \left( {{N_{{\rm{tx}}}} - 1} \right)\sin {\theta _{bxy}}}}} \right]}^{\rm{T}}}}}{{\sqrt {\left( {\kappa  + 1} \right){{\left\| {{\bf{d}}_b^{\rm{B}} - {{\bf{d}}_{xy}}} \right\|}^\xi }} }}, \label{eqsteering}
\end{align}
with ${\theta _{bxy}} = \arctan \left( {\left( {{{\left[ {{\bf{d}}_b^{\rm{B}}} \right]}_2} - {{\left[ {{{\bf{d}}_{xy}}} \right]}_2}} \right)/{{\left( {{{\left[ {{\bf{d}}_b^{\rm{B}}} \right]}_1} - {{\left[ {{{\bf{d}}_{xy}}} \right]}_1}} \right)}}} \right)$.

\subsubsection{Communication Channel}
Upon denoting  the position of communication user  $k$  by ${\bf{d}}_{k}^{\rm{C}} = {\left[ {x_{k}^{\rm{C}},y_{k}^{\rm{C}}} \right]^{\rm T}}$, the downlink channel response from BS $b$ to user $k$ on each subcarrier is
\begin{align}
{\bf{g}}_{bk}^{\rm{C}} =  {\bf{v}}\left( {{\bf{d}}_b^{\rm{B}},{\bf{d}}_k^{\rm{C}}} \right) + {\bf{g}}_{bk}^{{\rm{C}},{\rm{NL}}} \in {{\mathbb C}^{{N_{{\rm{tx}}}} \times 1}},
\end{align}
where ${\bf{v}}\left( {{\bf{d}}_b^{\rm{B}},{\bf{d}}_k^{\rm{C}}} \right)$ is   defined according to  \eqref{eqsteering}, while the NLoS component ${\bf{g}}_{bk}^{{\rm{C}},{\rm{NL}}}$  is independently distributed as a complex Gaussian vector with zero mean and covariance matrix
 ${\frac{1}{\kappa+1}}\left\| {{\bf{d}}_b^{\rm B} - {{\bf{d}}_{k}^{\rm C}}} \right\|^{-\xi}{\bf I}_{N_{\rm tx}}$ and zero mean.

\subsubsection{Radar Target Mobility}
We consider that $Q$ targets are separately located in different sub-regions and we denote the position and velocity of target $q$ by ${\bf{d}}_q^{\rm{R}} = {\left[ {x_q^{\rm{R}},y_q^{\rm{R}}} \right]^{\rm T}}$ and ${\bm{v}}_q^{\rm{R}} = \left[ v_q^{x},v_q^{y} \right]^{\rm T}$, respectively.
Then, the time delay and Doppler shift of the cascaded/round-trip channel from BS $b$  via target $q$ to  BS $b'$ are expressed as
  \begin{align}
{\tau _{bb'q}}& = \frac{{\left\| {{\bf{d}}_b^{\rm{B}} - {\bf{d}}_q^{\rm R}} \right\| + \left\| {{\bf{d}}_{b'}^{\rm{B}} - {\bf{d}}_q^{\rm R}} \right\|}}{c}, \label{eqq6}\\
{f_{bb'q}} &= \frac{{{{\left( {{\bm{v}}_q^{\rm{R}}} \right)}^{\rm T}}\left( {{\bf{d}}_b^{\rm{B}} - {\bf{d}}_q^{\rm R}} \right)}}{{\bar \lambda \left\| {{\bf{d}}_b^{\rm{B}} - {\bf{d}}_q^{\rm R}} \right\|}} + \frac{{{{\left( {{\bm{v}}_q^{\rm{R}}} \right)}^{\rm T}}\left( {{\bf{d}}_{b'}^{\rm{B}} - {\bf{d}}_q^{\rm R}} \right)}}{{\bar \lambda \left\| {{\bf{d}}_{b'}^{\rm{B}} - {\bf{d}}_q^{\rm R}} \right\|}},\label{eqq7}
\end{align}
respectively, where $c$ is the speed of light and $\bar \lambda$ is the wave length of the carrier frequency signal.

\section{Derivations of System Performances}
In this section, we derive the performance of sensing, communication, and radiation footprint.

\subsection{Radar Sensing Performance}
To sense the target position and velocity information, we enable the active BSs to transmit the ISAC signals in \eqref{eqisacsignal}.
Then, the received echo signals at active BS $b'$ from all targets after self-interference cancellation (SIC) \cite{xie2023bac,chen2023impact} are given by
\begin{align}
{\bf{r}}_{b'}^{\rm{R}}\left( t \right) =& \sum\limits_{b =1}^B {\sum\limits_{q = 1}^Q {{\zeta _{q}}{\bf{g}}_{b'q}^{\rm{R}}{{\left( {{\bf{g}}_{bq}^{\rm{R}}} \right)}^{\rm{H}}}{e^{{\rm{j}}2\pi {f_{bb'q}}t}}{{\bf{s}}_{b}}\left( {t - {\tau _{bb'q}}} \right)} }\nonumber\\
&+{\bf{n}}_{b'}^{\rm{R}}\left( t \right), {\rm for}\; 1\le b'\le B, \label{eq9}
\end{align}
where ${\zeta _{q}}$ is the radar cross-section coefficient of target $q$ and ${\bf{g}}_{bq}^{\rm{R}} = {\bf{v}}\left( {{\bf{d}}_b^{\rm{B}},{\bf{d}}_{q}^{\rm{R}}} \right)$ is the channel response between BS $b$ and target $q$.
 Here,  the impact of the NLoS component is ignored due to the significant attenuation in the cascaded/round-trip sensing channel among distributed BSs.
 Besides, ${\bf{n}}_{b'}^{\rm{R}}\left( t \right)$ is the received independent  Gaussian noise with power $M\Delta_fN_0$, where  $N_0$ is the noise power spectrum density.

Then, upon denoting the sampled signal ${\bf{r}}_{b'}^{\rm{R}}\left( t \right)$  at time ${t = lT + {T_{\rm cp}} + \frac{m}{M}{T_o}}$  by ${\bf{\tilde r}}_{b'ml}^{\rm{R}}$, we have
\begin{align}
{\bf{\tilde r}}_{b'ml}^{\rm{R}}
  &= \sum\nolimits_{b = 1}^B {\sum\nolimits_{q = 1}^Q {\left( {{\zeta _{q}}{\bf{g}}_{b'q}^{\rm{R}}{e^{{\rm{j}}2\pi {f_{bb'q}}lT}}} \right.} } \nonumber\\
 &\quad\times\left. {\sum\nolimits_{\tilde m = 0}^{M - 1} { {\bar s_{bq}^{ml}{e^{{\rm{j}}2\pi \tilde m\left( {\frac{m}{M} - {\Delta _f}{\tau _{bb'q}}} \right)}}} } } \right)+
{\bf{\tilde n}}_{b'ml}^{\rm{R}},
 \end{align}
where $\bar s_{bq'}^{ml} = {\left( {{\bf{g}}_{bq'}^{\rm{R}}} \right)^{\rm H}}{\bf{\hat s}}_b^{ml}$ with ${\bf{\hat s}}_b^{ml}$ defined in \eqref{eq2} and ${\bf{\tilde n}}_{b'ml}^{\rm{R}}$ is the corresponding sampled noise term with power $M\Delta_fN_0$.

Next, we utilize the zero-forcing (ZF) receive beamforming ${\bf w}_{bq}^{\rm zf}$ with the unit power
for target $q$ to mitigate the interference of other targets, i.e., ${\bf w}_{bq}^{\rm zf}$ is the normalized $q$-th column of matrix ${{\bf{G}}_b}{\left( {{\bf{G}}_b^{\rm{H}}{{\bf{G}}_b}} \right)^{ - 1}}$ with unit power where ${{\bf{G}}_b} = \left[ {{\bf{g}}_{b1}^{\rm{R}},\dots,{\bf{g}}_{bQ}^{\rm{R}}} \right]$.
Then, we  have
  \begin{align}
\bar r_{b'q}^{ml} &= \frac{1}{M}\sum\nolimits_{m' = 0}^{M - 1} {{{\left( {{\bf{w}}_{b'q}^{\rm zf}} \right)}^{\rm H}}{\bf{\tilde r}}_{b'm'l}^{\rm{R}}} {e^{{\rm{ - j}}2\pi m'\frac{m}{M}}}\nonumber\\
& = \sum\nolimits_{b = 1}^B {\underbrace {{{\bar \zeta }_{b'q}}\bar s_{bq}^{ml}{e^{{\rm{j}}2\pi \left( {{f_{bb'q}}lT - m{\Delta _f}{\tau _{bb'q}}} \right)}}}_{\mu _{bb'q}^{ml}}}  + \bar n_{b'q}^{{\rm{R}},ml},
\end{align}
where ${{\bar \zeta }_{b'q}} = {\zeta _{q}}{\left( {{\bf{w}}_{b'q}^{\rm zf}} \right)^{\rm H}}{\bf{g}}_{b'q}^{\rm{R}}$, while  ${\tau _{bb'q}}$ and $ {f_{bb'q}}$ are defined in \eqref{eqq6} and \eqref{eqq7}, respectively.
Also, $\bar n_{b'q}^{{\rm{R}}, ml}=\frac{1}{M}\sum\nolimits_{m' = 0}^{M - 1} {{{\left( {{\bf{w}}_{b'q}^{{\rm{zf}}}} \right)}^{\rm{H}}}{\bf{\tilde n}}_{b'm'l}^{\rm{R}}} {e^{{\rm{ - j}}2\pi m'\frac{m}{M}}}$ is the equivalent Gaussian noise with mean zero and covariance $\sigma=\Delta_fN_0$  due to the unit power of  ${\bf w}_{bq}^{\rm zf}$.

To obtain the CRBs of the sensing parameters of target $q$, i.e., ${{\bm{\eta }}_q} = {\left[ {x_q^{\rm{R}},y_q^{\rm{R}},v_q^x,v_q^y} \right]^{\rm{T}}}$, we first calculate the Fisher's information matrix (FIM) respect to ${{\bm{\eta }}_q}$ \cite{van2004detection}.
Specifically, upon denoting the received signals for sensing target $q$ at the CPU in a vector form by ${{{\bf{\bar r}}}_q} = {\left[ {{\bf{\bar r}}_{1q}^{\rm{T}},{\bf{\bar r}}_{2q}^{\rm{T}}, \ldots ,{\bf{\bar r}}_{Bq}^{\rm{T}}} \right]^{\rm{T}}}$ with ${{{\bf{\bar r}}}_{b'q}} = {\left[ {\bar r_{b'q}^{00},\bar r_{b'q}^{01}, \ldots ,\bar r_{b'q}^{(M - 1)(L - 1)}} \right]^{\rm{T}}}$ for $1\le b'\le B$, and the moving parameters of target $q$ by ${{\bm{\omega }}_q} ={[\tau _{11q}, \ldots ,\tau _{BBq},{f_{11q}}, \ldots ,{f_{BBq}}]^{\rm{T}}}$, the  FIM of ${{\bm{\eta }}_q}$ is
 \begin{align}{{\bf{J}}_q}{\rm{ = }}\frac{{\partial {{\bm{\omega }}_q}}}{{\partial {\bm \eta _q}}}\left[ {\begin{array}{*{20}{c}}
{{{\bf{J}}_q^{\tau \tau }}}&{{{\bf{J}}_q^{\tau f}}}\\
{{{\bf{J}}_q^{\tau f}}}&{{{\bf{J}}_q^{ff}}}
\end{array}} \right]{\left( {\frac{{\partial {{\bm{\omega }}_q}}}{{\partial {\bm \eta _q}}}} \right)^{\rm{T}}}\in{\mathbb C}^{4\times4},\label{eq11}
 \end{align}
 where
$\frac{{\partial {{\bm{\omega }}_q}}}{{\partial {{\bm \eta} _q}}} $
  and the $(i,j)$-th elements of $ {\bf{J}}_q^{\tau \tau }$,  $ {\bf{J}}_q^{\tau f }$, and $ {\bf{J}}_q^{ff }$ with $ i = \left( {b - 1} \right)B + b'$ and $j = \left( {\tilde b- 1} \right)B + \tilde b' $ for $1\le b, b',\tilde b, \tilde b' \le B$ are, respectively, given by
 \begin{align}
 &\!\!\!\frac{{\partial {{\bm{\omega }}_q}}}{{\partial {{\bm\eta} _q}}} \!= \!\left[ {\begin{array}{*{20}{c}}
{\frac{{\partial {\tau _{11q}}}}{{\partial x_q^{\rm{R}}}}}& \ldots &{\frac{{\partial {\tau _{BBq}}}}{{\partial x_q^{\rm{R}}}}}&{\frac{{\partial {f_{11q}}}}{{\partial x_q^{\rm{R}}}}}& \ldots &{\frac{{\partial {f_{BBq}}}}{{\partial x_q^{\rm{R}}}}}\\
{\frac{{\partial {\tau _{11q}}}}{{\partial y_q^{\rm{R}}}}}& \ldots &{\frac{{\partial {\tau _{BBq}}}}{{\partial y_q^{\rm{R}}}}}&{\frac{{\partial {f_{11q}}}}{{\partial y_q^{\rm{R}}}}}& \ldots &{\frac{{\partial {f_{BBq}}}}{{\partial y_q^{\rm{R}}}}}\\
0& \ldots &0&{\frac{{\partial {f_{11q}}}}{{\partial v_q^x}}}& \ldots &{\frac{{\partial {f_{BBq}}}}{{\partial v_q^x}}}\\
0& \ldots &0&{\frac{{\partial {f_{11q}}}}{{\partial v_q^y}}}& \ldots &{\frac{{\partial {f_{BBq}}}}{{\partial v_q^y}}}
\end{array}} \right],\\
&\!\!\!{\left[ {{\bf{J}}_q^{xy}} \right]_{ij}} \!=\! {\mathbb{E}}\left[- {\frac{{{\partial ^2}\log { P}\left( {{{{\bf{\bar r}}}_q}\left| {{{\bm{\omega }}_q}} \right.} \right)}}{{\partial {x_{bb'q}}\partial {y_{\tilde b\tilde b'q}}}}} \right] \;{\rm with }\;x,y\in \left\{\tau,f\right\},\\
&\!\!\!P({{{{\bf{\bar r}}}_q}\left| {{{\bm \omega}_q}}\right.}) \!\propto\! \exp \!\left( \!\!{\sum\limits_{b' = 1}^B {\sum\limits_{m = 0}^{M - 1} {\sum\limits_{l = 0}^{L - 1} {\frac{a_{b'}{|\bar r_{b'q}^{ml} - \sum\limits_{b = 1}^B {\mu _{bb'q}^{ml}} {|^2}}}{\sigma }} } } }\!\! \right).
\end{align}

\begin{theorem}\label{theorem1}
Ignoring the coupling effects between distance and velocity \cite{zhong2023performance}, the FIM ${{\bf{J}}_q}$ can be approximated by
\begin{align}
{{\hat { \bf J}}_q} = \left[ {\begin{array}{*{20}{c}}
{\sum\nolimits_{b = 1}^B {\lambda _{bq}^{\rm{R}}{\bf{H}}_{bq}^{\rm{D}}} }&{\bf 0}\\
{\bf 0}&{\sum\nolimits_{b = 1}^B {\lambda _{bq}^{\rm{R}}{\bf{H}}_{bq}^{\rm{V}}} }
\end{array}} \right],\label{eq17}
\end{align}
where ${\bf{H}}_{bq}^{\rm{D}} $ is defined in \eqref{eqHbq26} at the top of next page, and
\begin{align}
{\bf{H}}_{bq}^{\rm{V}}& = \sum\limits_{b' = 1}^B {\left[ {\begin{array}{*{20}{c}}
{{{\left( {\frac{{\partial {f_{bb'q}}}}{{\partial v_q^x}}} \right)}^2}\beta _{b'q}^{ff}}&{\frac{{\partial {f_{bb'q}}}}{{\partial v_q^x}}\frac{{\partial {f_{bb'q}}}}{{\partial v_q^y}}\beta _{b'q}^{ff}}\\
{\frac{{\partial {f_{bb'q}}}}{{\partial v_q^x}}\frac{{\partial {f_{bb'q}}}}{{\partial v_q^y}}\beta _{b'q}^{ff}}&{{{\left( {\frac{{\partial {f_{bb'q}}}}{{\partial v_q^y}}} \right)}^2}\beta _{b'q}^{ff}}
\end{array}} \right]} ,\label{eqp15}\\
{\lambda _{bq}^{\rm R}}& ={a_b}\left( {\sum\nolimits_{k = 1}^K {{{\left| {{{\left( {{\bf{g}}_{bq}^{\rm{R}}} \right)}^{\rm{H}}}{\bf{w}}_{bk}^{\rm{C}}} \right|}^2}}  + {{\left\| {{{\left( {{\bf{g}}_{bq}^{\rm{R}}} \right)}^{\rm{H}}}{\bf{W}}_b^{\rm{R}}} \right\|}^2}} \right),\label{eqp16}
\end{align}
\begin{align}
\beta _{b'q}^{\tau \tau }& = \frac{{4\left( {M - 1} \right)M\left( {2M - 1} \right)L}}{{3{\sigma }}}{\pi ^2}\Delta _f^2\bar \xi _{b'q}^2,\label{eqp17}\\
\beta _{b'q}^{ff}& = \frac{{4\left( {L - 1} \right)L\left( {2L - 1} \right)M}}{{3{\sigma }}}{\pi ^2}{T^2}\bar \xi _{b'q}^2,\label{eqp18}\\
\beta _{b'q}^{\tau f} &= \frac{{2ML\left( {M - 1} \right)\left( {L - 1} \right)}}{{{\sigma}}}{\pi ^2}T{\Delta _f}\bar \xi _{b'q}^2.\label{eqp19}
\end{align}
\end{theorem}
\begin{IEEEproof}
Please refer to Appendix \ref{appendix1}.
\end{IEEEproof}

\begin{figure*}[ht]
\begin{subequations} \label{eqHbq26}
\begin{align}
&{\left[ {{\bf{H}}_{bq}^{\rm{D}}} \right]_{1,1}} = \sum\nolimits_{b' = 1}^B {\left( {{{\left( {\frac{{\partial {\tau _{bb'q}}}}{{\partial x_q^{\rm{R}}}}} \right)}^2}\beta _{b'q}^{\tau \tau } + {{\left( {\frac{{\partial {f_{bb'q}}}}{{\partial x_q^{\rm{R}}}}} \right)}^2}\beta _{b'q}^{ff} -2 \frac{{\partial {f_{bb'q}}}}{{\partial x_q^{\rm{R}}}}\frac{{\partial {\tau _{bb'q}}}}{{\partial x_q^{\rm{R}}}}\beta _{b'q}^{\tau f}} \right)}, \\
&{\left[ {{\bf{H}}_{bq}^{\rm{D}}} \right]_{1,2}} = {\left[ {{\bf{B}}_b^1} \right]_{2,1}} = \sum\nolimits_{b' = 1}^B {\left( {\frac{{\partial {\tau _{bb'q}}}}{{\partial x_q^{\rm{R}}}}\frac{{\partial {\tau _{bb'q}}}}{{\partial y_q^{\rm{R}}}}\beta _{b'q}^{\tau \tau } + \frac{{\partial {f_{bb'q}}}}{{\partial x_q^{\rm{R}}}}\frac{{\partial {f_{bb'q}}}}{{\partial y_q^{\rm{R}}}}\beta _{b'q}^{ff} - \left( {\frac{{\partial {f_{bb'q}}}}{{\partial x_q^{\rm{R}}}}\frac{{\partial {\tau _{bb'q}}}}{{\partial y_q^{\rm{R}}}} + \frac{{\partial {f_{bb'q}}}}{{\partial y_q^{\rm{R}}}}\frac{{\partial {\tau _{bb'q}}}}{{\partial x_q^{\rm{R}}}}} \right)\beta _{b'q}^{\tau f}} \right)},  \\
&{\left[ {{\bf{H}}_{bq}^{\rm{D}}} \right]_{2,2}} = \sum\nolimits_{b' = 1}^B {\left( {{{\left( {\frac{{\partial {\tau _{bb'q}}}}{{\partial y_q^{\rm{R}}}}} \right)}^2}\beta _{b'q}^{\tau \tau } + {{\left( {\frac{{\partial {f_{bb'q}}}}{{\partial y_q^{\rm{R}}}}} \right)}^2}\beta _{b'q}^{ff} - 2\frac{{\partial {f_{bb'q}}}}{{\partial y_q^{\rm{R}}}}\frac{{\partial {\tau _{bb'q}}}}{{\partial y_q^{\rm{R}}}}\beta _{b'q}^{\tau f}} \right)},
\end{align}
\end{subequations}
\hrulefill
\end{figure*}

From \cite{van2004detection}, we know the CRBs of the position and velocity parameters in ${{\bm{\eta }}_q} = {\left[ {x_q^{\rm{R}},y_q^{\rm{R}},v_q^x,v_q^y} \right]^{\rm{T}}}$ are given by the diagonal components of matrix ${{\bf{\hat J}}_q^{ - 1}}$.
Then, we define the sensing accuracy as the CRBs of position and velocity parameters, which include both the x- and y-axis components, i.e.,
\begin{align}
\varepsilon _q^{\rm D} =\max \left( {{{\left[ {\hat{\bf{ J}}_q^{ - 1}} \right]}_{1,1}},{{\left[ {{\hat{  \bf J}}_q^{ - 1}} \right]}_{2,2}}} \right),\label{eq0241}\\
\varepsilon _q^{\rm V} = \max \left( {{{\left[ {\hat {\bf{ J}}_q^{ - 1}} \right]}_{3,3}},{{\left[ {{\hat {  \bf J}}_q^{ - 1}} \right]}_{4,4}}} \right).\label{eq0242}
\end{align}

\subsection{Communication Performance}
Upon transmitting the ISAC signals in \eqref{eqisacsignal}, the received $l$-th symbol  on subcarrier $m$ at user $k$ is denoted by
 \begin{align}
 r_{kml}^{\rm{C}} = \sum\limits_{b = 1}^B  a_b{{{\left( {{\bf{g}}_{bk}^{\rm{C}}} \right)}^{\rm{H}}}}
\left( \sum\limits_{k = 1}^K {{\bf{w}}_{bk}^{\rm{C}}\hat s_{kml}^{\rm{C}} + {\bf{W}}_{b}^{\rm{R}}{\bf{\hat s}}_{bml}^{\rm{R}}}\right) + n_{kml}^{\rm{C}},\label{eq025}
  \end{align}
where $n_{kml}^{\rm C}$ is the received complex Gaussian noise with zero mean and covariance $\sigma_k=\Delta_fN_0$.
Then, since the sensing signals are predetermined sequences known at the transceivers, they can be canceled by the communication users. Therefore,  the transmission rate (bits/s/Hz) for user $k$  is given  by
\begin{align}
{C_k}\! =\! {\log _2}\left( {1 \!+\! \frac{{{{\left| {\sum\nolimits_{b = 1}^B {{a_b}{{\left( {{\bf{g}}_{bk}^{\rm{C}}} \right)}^{\rm{H}}}{\bf{w}}_{bk}^{\rm{C}}} } \right|}^2}}}{{\sum\nolimits_{k' \ne k}^K {{{\left| {\sum\nolimits_{b = 1}^B {{a_b}{{\left( {{\bf{g}}_{bk}^{\rm{C}}} \right)}^{\rm{H}}}{\bf{w}}_{bk'}^{\rm{C}}} } \right|}^2}}  \!+\! {\sigma _k}}}} \right).\label{eq26}
\end{align}

\subsection{Radiation Footprint Evaluation}

To suppress potential interference to other networks sharing the same spectrum, the studied ISAC network is enabled to autonomously control its radiation footprint across the entire service area.
To do so, we aim to evaluate and control the BS radiation footprint on the sub-regions without accommodating any communication users and targets, i.e., ${\cal S}_o={\cal S}\backslash {{\cal S}_{\rm RC}}$, where ${\cal S}_{\rm RC}$ represents the index set of sub-regions containing at least one user or target:
\begin{align}
{S_{{\rm{RC}}}} = \left\{ {\left( {x,y} \right)\left| \begin{array}{l}
\left( {\left\lceil {\frac{{x_k^{\rm{C}}}}{{{\Delta _x}}}} \right\rceil ,\left\lceil {\frac{{y_k^{\rm{C}}}}{{{\Delta _y}}}} \right\rceil } \right),1 \le k \le K, \\
\left( {\left\lceil {\frac{{x_q^{\rm{R}}}}{{{\Delta _x}}}} \right\rceil ,\left\lceil {\frac{{y_q^{\rm{R}}}}{{{\Delta _y}}}} \right\rceil } \right),1 \le q \le Q
\end{array} \right.} \right\}.
  \end{align}

Upon transmitting the  signals in \eqref{eqisacsignal} with  the prorogation channel response in  \eqref{eqaa3}, the signal propagated in sub-region $(x,y)\in{\cal S}_o$  on subcarrier $m$ is calculated as
 \begin{align}
 r_{xy}^{\rm{I}}\!=\! \sum\limits_{b = 1}^B {\!\!{a_b}{{\left( {{\bf{g}}_{bxy}^{\rm{L}} \!+\! {\bf{g}}_{bxy}^{{\rm{NL}}}} \right)}^{\rm{H}}}}
 \!\!\left( \sum\limits_{k = 1}^K {{\bf{w}}_{bk}^{\rm{C}}\hat s_{kml}^{\rm{C}}\! +\! {\bf{W}}_{b}^{\rm{R}}{\bf{\hat s}}_{bml}^{\rm{R}}}\right).
  \end{align}

However, evaluating the exact radiation level of $r_{xy}^{\rm{I}} $
requires estimating the instantaneous CSI of the propagation channel in \eqref{eqaa3} for each sub-region in each transmission frame.
This is quite challenging,  as it necessitates the deployment of numerous sensor devices and introduces significant estimation overhead.
Therefore,  this paper focuses on the expectation value of the signal radiation power.
Specifically, this average value depends solely on the statistical information of the propagation channel, including the angle and distance associated with each sub-region, as well as the Rician factor.
Besides, these parameters can be pre-estimated during the system configuration stage, thus eliminating the need of CSI estimation for every sub-region in each transmission block.
Mathematically, in sub-region $(x,y)\in{\cal S}_o$, the expected radiation power on each subcarrier can be expressed as
   \begin{align}
{I_{xy}} = &{\mathbb{E}}\left( {{{\left| {r_{xy}^{\rm{I}}} \right|}^2}} \right)\nonumber\\
   =&\sum\limits_{k = 1}^K {{{\left| {\sum\limits_{b = 1}^B {{a_b}} {{\left( {{\bf{g}}_{bxy}^{\rm{L}}} \right)}^{\rm{H}}}{\bf{w}}_{bk}^{\rm{C}}} \right|}^2}}  + \sum\limits_{b = 1}^B {{a_b}} {\left\| {{{\left( {{\bf{g}}_{bxy}^{\rm{L}}} \right)}^{\rm{H}}}{\bf{W}}_b^{\rm{R}}} \right\|^2}\nonumber\\
& + \sum\limits_{b = 1}^B {{a_b}} {{\bar \kappa }_{bxy}}\left( {{{\left\| {{\bf{W}}_b^{\rm{R}}} \right\|}^2} + \sum\limits_{k = 1}^K {{{\left\| {{\bf{w}}_{bk}^{\rm{C}}} \right\|}^2}} } \right),
  \end{align}
where  ${{\bf{g}}_{bxy}^{\rm{L}}} $ and $\bar\kappa _{bxy}$ are  static and defined after \eqref{eqaa3}.

\section{Problem Formulation and Transformation}
This section formulates the joint BS activation and beamforming coordination problem and transforms it into an equivalent but tractable form for developing efficient algorithms.

\subsection{Problem Formulation}
To balance the performances between S\&C services, we propose the following weighted normalized communication and radar sensing utility functions, i.e.,
\begin{align}
{U_{\rm{C}}} &= \sum\nolimits_{k = 1}^K w_k^{\rm{C}}{\frac{{{C_k}}}{{C_k^{\min }}}}, \label{eqap30}\\
{U_{\rm{R}}} &=\sum\nolimits_{q = 1}^Q {w_q^{\rm{R}}\left( {1 - \max \left( {\frac{{\varepsilon _q^{\rm{D}}}}{{\varepsilon _q^{{\rm{D}},\max }}},\frac{{\varepsilon _q^{\rm{V}}}}{{\varepsilon _q^{{\rm{V}},\max }}}} \right)} \right)} ,\label{eqap31}
\end{align}
where  ${\varepsilon _q^{{\rm{D}},\max }}$,  ${\varepsilon _q^{{\rm{V}},\max }}$, and ${C_k}$ are defined in \eqref{eq0241}, \eqref{eq0242}, \eqref{eq26}, respectively;
 ${C_k^{\min }}$ is the minimal transmission rate requirement for user $q$;
  ${\varepsilon _q^{{\rm D},\max }}$ and ${\varepsilon _q^{{\rm V},\max }}$ are the maximum sensing errors of the position and velocity of target $q$, respectively;
  $w_k^{\rm C}$ and $w_q^{\rm R}$ are positive constant weighting parameters used to balance the trade-offs among sensing targets and communication users, respectively.

Then, we attempt to concurrently maximize the S\&C utilities while minimizing the operational costs of all BSs.
Hence, we adopt the proportional trade-off metric  \cite{vielhaus2025evaluating,panayiotou2023balancing} and formulate the joint BS activation and beamforming coordination optimization problem using the proportional weighted cost-efficient S\&C utility objective, i.e.,
\begin{align}\mathop {\max }\limits_{{\bf{a}},{\bf{W}} } &\;U \buildrel \Delta \over =\frac{{U_{\rm{C}}^{{\alpha _{\rm{C}}}}U_{\rm{R}}^{{\alpha _{\rm{R}}}}}}{{{{\left( {\sum\nolimits_{b = 1}^B {{a_b}{o_b}} } \right)}^{{\alpha _{\rm{A}}}}}}}\tag{\bf P0} \label{eqp1}\\
{\rm{s}}.{\rm{t}}.\;&{C_k} \ge C_k^{\min }, \forall k,\tag{\rm C1}\label{cons1}\\
& \varepsilon _q^{\rm{D}} \le \varepsilon _q^{{\rm{D}},{\rm{max}}}, \forall q,\tag{\rm C2}\label{cons2}\\
&\varepsilon _q^{\rm{V}} \le \varepsilon _q^{{\rm{V}},{\rm{max}}}, \forall q,\tag{\rm C3}\label{cons3}\\
&M{I_{xy}} \le  I_{xy}^{{\rm{max}}}  \tag{\rm C4}, \forall (x,y)\in{\cal S}_o,\label{cons4}\\
&{{{\sum\nolimits_{k = 1}^K {\left\| {{\bf{w}}_{bk}^{\rm{C}}} \right\|^2} }} + {{\left\| {{\bf{W}}_b^{\rm{R}}} \right\|}^2}}\le {a_bp_{b }^{\max}}, \forall b,\tag{\rm C5}\label{cons5}\\
&a_b \in\left\{0,1\right\}, \forall b,\tag{\rm C6}\label{cons6}\\
&1\le\sum\nolimits_{b = 1}^B {{a_b}}  \le {N_{{\rm{bs}}}},\tag{\rm C7}\label{cons7}
\end{align}
where ${\bf a}=\left[a_1,a_2,\ldots,a_B\right]$ denotes the set of optimization variables including all BS activation scheduling statuses, and
 ${\bf W}$ denotes the set of optimization variables comprising all beamforming vectors  ${\bf{w}}_{bk}^{\rm{C}}$ and  matrices ${\bf{W}}_{b}^{\rm{R}} $ for $ \forall b$ and $ \forall k$.
Here, the positive constant weights ${{\alpha_{\rm{C}}}}$, ${{\alpha_{\rm{R}}}}$, and ${{\alpha_{\rm{A}}}}$ are utilized to achieve various trade-offs in communication performance, sensing performance, and system operation costs, respectively.
Additionally, the positive constant parameter ${o_b\ge1}$  is used to characterize the normalized fixed operation cost of BS  $b$.
Moreover, \eqref{cons1} specifies  the  minimal rate requirement $C_k^{\min }$  user $k$;
\eqref{cons2} and \eqref{cons3} impose the maximum sensing error tolerance  $\varepsilon _q^{{\rm{D}},{\rm{max}}}$ and $\varepsilon _q^{{\rm{V}},{\rm{max}}}$
for the position and velocity of target $q$;
\eqref{cons4} implies that the signal radiation power (i.e., the potential interference to other networks) at sub-region $(x,y)$ across all sub-carriers should be lower than threshold $ I_{xy}^{{\rm{max}}}$; \eqref{cons5} denotes that the transmission power of BS $b$ should be smaller than $p_b^{\rm max}$ if it is active; \eqref{cons6}  denotes the binary activation-deactivation status of BS $b$;  \eqref{cons7}  denotes the number of active BSs should be greater than one and less than the constant threshold ${N_{\text{bs}}}$.

\subsection{Problem Transformation}
Problem  \ref{eqp1} is challenging to solve since it falls within the MINLP family \cite{chen2023digital,jia2023new}.
To develop efficient algorithms, we transform \ref{eqp1} into an equivalent but tractable formulation.

Firstly, we define ${\bf{V}} = \left[ {{\bf{V}}_{1,}^{\rm{R}},\ldots,{\bf{V}}_B^{\rm{R}},{\bf{V}}_1^{\rm{C}},\ldots,{\bf{V}}_K^{\rm{C}}} \right]$, where  ${\bf{V}}_b^{\rm{R}} =
a_b{\bf{W}}_b^{\rm{R}}{\left( {{\bf{W}}_b^{\rm{R}}} \right)^{\rm{H}}}$ and ${\bf{V}}_k^{\rm{C}} = {\bf{w}}_k^{\rm{C}}{\left( {{\bf{w}}_k^{\rm{C}}} \right)^{\rm{H}}} $ with ${\bf{w}}_k^{\rm{C}} = {[ {{{(a_1{{\bf{w}}_{1k}^{\rm{C}}} )}^{\rm{T}}},\ldots,{{(a_B {{\bf{w}}_{Bk}^{\rm{C}}} )}^{\rm{T}}}} ]^{\rm{T}}} $.
Then, we define
${\bf{\hat G}}_{bk}^{\rm{C}} = {\bf{g}}_{bk}^{\rm{C}}{\left( {{\bf{g}}_{bk}^{\rm{C}}} \right)^{\rm{H}}}$ and
${\bf{G}}_k^{\rm{C}} = {\bf{g}}_k^{\rm{C}}{\left( {{\bf{g}}_k^{\rm{C}}} \right)^{\rm{H}}}$ with  ${\bf{g}}_k^{\rm{C}} = [ {{{( {{\bf{g}}_{1k}^{\rm{C}}} )}^{\rm{T}}},\ldots,{{( {{\bf{g}}_{Bk}^{\rm{C}}} )}^{\rm{T}}}} ]^{\rm{T}} $.
Upon introducing auxiliary variables ${{\bf z} ^{\rm C}} =  \left[ {z _1^{\rm{C}},\ldots,z _K^{\rm{C}}} \right]$, the communication utility defined in \eqref{eqap30}  is reformulated as:
\begin{align}{U_{\rm{C}}} =& \mathop {\sup }\limits_{{{\bf{z}}^{\rm{C}}}} \left\{ {{U_{\rm{C}}}\left( {{{\bf{z}}^{\rm{C}}}} \right)  \buildrel \Delta \over =  \sum\nolimits_{k = 1}^K {\frac{{w_k^{\rm{C}}}}{{C_k^{\min }\ln2}}  z_k^{\rm{C}}} } \right\}\label{eq330}\\
&{\rm{s}}{\rm{.t.}}\; z_k^{\rm{C}} \le \ln \left( {1 + \frac{{{{\cal F}_k}\left( {\bf{V}} \right)}}{{{{\cal H}_k}\left( {\bf{V}} \right)}}} \right)\buildrel \Delta \over = {\cal C}_k \left( {\bf{V}} \right),\forall k,
 \tag{\rm C8}\label{cons8}\\
&\quad\;\; z _k^{\rm{C}} \ge { {C_k^{\min }}} \ln2 ,\forall k,\tag{\rm C9} \label{cons9}\\
&\quad\;\;  {\rm{rank}}\left( {{\bf{V}}_k^{\rm{C}}} \right) \le 1,\forall k,\tag{\rm C10} \label{cons10}\\
&\quad\;\;  {\cal P}_b\left({\bf V}\right)  \le {a_b}p_b^{\max },\forall b,\tag{\rm C11}\label{cons11}
\end{align}
where \eqref{cons8} and \eqref{cons10} ensure the transformation equivalence. Besides, \eqref{cons9} and \eqref{cons11} are due to \eqref{cons1} and \eqref{cons5}, respectively.
Here, we denote
\begin{align}
{\cal P}_b\left({\bf V}\right)&=\sum\nolimits_{k = 1}^K {{\rm{Tr}}\left( {{{\bf{B}}_b}{\bf{V}}_k^{\rm{C}}} \right) + {\rm{Tr}}\left( {{\bf{V}}_b^{\rm{R}}} \right)}\\
{\cal F}_k \left( {\bf{V}} \right)& = {\left| {\sum\nolimits_{b = 1}^B a_b{{{\left( {{\bf{g}}_{bk}^{\rm{C}}} \right)}^{\rm{H}}}{\bf{w}}_{bk}^{\rm{C}}} } \right|^2} = {\rm{Tr}}\left( {{\bf{G}}_k^{\rm{C}}{\bf{V}}_k^{\rm{C}}} \right),\label{eq33a} \\
 {\cal H}_k\left( {\bf{V}} \right) & = \sum\nolimits_{k' \ne k}^K {{{\left| {\sum\nolimits_{b = 1}^B {{a_b}{{\left( {{\bf{g}}_{bk}^{\rm{C}}} \right)}^{\rm{H}}}{\bf{w}}_{bk'}^{\rm{C}}} } \right|}^2}}  + {\sigma _k}\nonumber\\
 &= \sum\nolimits_{k' \ne k}^K {{\rm{Tr}}\left( {{\bf{G}}_k^{\rm{C}}{\bf{V}}_{k'}^{\rm{C}}} \right)}  + {\sigma _k},  \label{eq34a} \\
{{\bf{B}}_b} &= {\rm{diag(}}\underbrace {0, \ldots ,0}_{\left( {b - 1} \right){N_{{\rm{tx}}}}},\underbrace {1, \ldots ,1}_{{N_{{\rm{tx}}}}},\underbrace {0, \ldots ,0}_{\left( {B - b} \right){N_{{\rm{tx}}}}}{\rm{)}}.
\end{align}

Similarly, upon introducing auxiliary variables ${{  z} ^{\rm R}}$ and ${\bm \epsilon}  = \left[ {\epsilon _1^{\rm{R}},\ldots,\epsilon _Q^{\rm{R}},\epsilon _1^{\rm{D}},\ldots,\epsilon _Q^{\rm{D}},\epsilon _1^{\rm{V}},\ldots,\epsilon _Q^{\rm{V}}} \right] $,  the   sensing utility defined in \eqref{eqap31} is reformulated as:
\begin{align}
{U_{\rm{R}}}=  &\mathop {\sup }\limits_{{z^{\rm{R}}},{\bm \epsilon} }  {{z^{\rm{R}}}} \label{cons370}\\
{\rm{s}}{\rm{.t}}{\rm{.}}\;
&\epsilon _q^{\rm{R}} \le \min \left( {1 - \frac{{\epsilon _q^{\rm{D}}}}{{\varepsilon _q^{{\rm{D}},\max }}},1 - \frac{{\epsilon _q^{\rm{V}}}}{{\varepsilon _q^{{\rm{V}},\max }}}} \right),\forall q,\tag{\rm C12} \label{cons12}\\
&\epsilon _q^{\rm{D}} \le \epsilon _q^{{\rm{D}},\max },\epsilon _q^{\rm{V}} \le \varepsilon _q^{{\rm{V}},\max },\forall q,\tag{\rm C13} \label{cons13}\\
&{z^{\rm{R}}} \le \sum\nolimits_{q = 1}^Q {w_q^{\rm{R}}\epsilon _q^{\rm{R}}} , \tag{\rm C14} \label{cons14}\\
&\max \left( {{{\left[ {{\bf{J}}_q^{ - 1}} \right]}_{1,1}},{{\left[ {{\bf{J}}_q^{ - 1}} \right]}_{2,2}}} \right) \le \epsilon _q^{\rm{D}},\forall q,\tag{\rm C15} \label{cons15}\\
&\max \left( {{{\left[ {{\bf{J}}_q^{ - 1}} \right]}_{3,3}},{{\left[ {{\bf{J}}_q^{ - 1}} \right]}_{4,4}}} \right) \le \epsilon _q^{\rm{V}},\forall q, \tag{\rm C16} \label{cons16}
 \end{align}
where \eqref{cons12} - \eqref{cons16} are due to the sensing utility reformulation with \eqref{cons2}, and \eqref{cons3}.

Given ${\bf{\hat G}}_{bq}^{\rm{R}} = {\bf{g}}_{bq}^{\rm{R}}{\left( {{\bf{g}}_{bq}^{\rm{R}}} \right)^{\rm{H}}}$ and
${\bf{G}}_q^{\rm{R}} = {\bf{g}}_q^{\rm{R}}{\left( {{\bf{g}}_q^{\rm{R}}} \right)^{\rm{H}}}$ with  ${\bf{g}}_q^{\rm{R}} = [ {{{( {{\bf{g}}_{1q}^{\rm{C}}} )}^{\rm{T}}},\ldots,{{( {{\bf{g}}_{Bq}^{\rm{R}}} )}^{\rm{T}}}} ]^{\rm{T}} $, constraints \eqref{cons15} and  \eqref{cons16} can be equivalently replaced by following convex constraints\footnote{
By defining the following equations,
\begin{align}
{\bf{X}}^{-1} = {\left[ {\begin{array}{*{20}{c}}
{{x_{11}}}&{{x_{12}}}\\
{{x_{12}}}&{{x_{22}}}
\end{array}} \right]^{ - 1}} = \frac{1}{{{x_{11}}{x_{22}} - x_{12}^2}}\left[ {\begin{array}{*{20}{c}}
{{x_{22}}}&{{x_{12}}}\\
{{x_{12}}}&{{x_{11}}}
\end{array}} \right], \nonumber
\end{align}
we know ${\left[ {\bf{X}} \right]_{jj}^{-1}} \le \varepsilon $ holds if $
\frac{{{x_{ll}}}}{{{x_{jj}}{x_{ll}} - x_{12}^2}} \le \varepsilon
 \Leftrightarrow \frac{1}{\varepsilon } + \frac{{x_{12}^2}}{{{x_{ll}}}} \le {x_{jj}}$ for $j,l \in \left\{ {1,2} \right\}$   and $j \ne l$ Besides, $\frac{{x_{12}^2}}{{{x_{ll}}}}$ is jointly convex  with respect to $x_{12}$ and $x_{ll}$  when $x_{ll}>0$.}:
\begin{align}
& \frac{1}{{\epsilon _q^{\rm{X}}}} + \frac{{{{\left( {{ {\cal J}}_{q12}^{\rm{X}}\left( {\bf{V}} \right)} \right)}^2}}}{{{ {\cal J}}_{qjj}^{\rm{X}}\left( {\bf{V}} \right)}} \le { {\cal J}}_{qll}^{\rm{X}}\left( {\bf{V}} \right), \nonumber\\
 &\qquad{\rm for} \;  {\rm{X}} \in \left\{ {{\rm{D}},{\rm{V}}} \right\}, j,l \in \left\{ {1,2} \right\},\;j \ne l,\;  {\rm and}\; \forall q, \tag{\rm C17} \label{cons17}
\end{align}
where
\begin{align}
{ {\cal J}}_{qjl}^{\rm{X}}\left( {\bf{V}} \right)\!& =\! \sum\nolimits_{b = 1}^B {\lambda _{bq}^{\rm{R}}\left( {\bf{V}} \right){{\left[ {{\bf{H}}_{bq}^{\rm{X}}} \right]}_{jl}}}\\
\lambda _{bq}^{\rm{R}}\left( {\bf{V}} \right)\! &=\!  {\sum\nolimits_{k = 1}^K {{\rm{Tr}}\left( {{{\bf{B}}_b}{\bf{G}}_q^{\rm{R}}{{\bf{B}}_b}{\bf{V}}_k^{\rm{C}}} \right)}  + {\rm{Tr}}\left( {{\bf{\hat G}}_{bq}^{\rm{R}}{\bf{V}}_b^{\rm{R}}} \right)}. \label{eqa38}
 \end{align}

Next, constraint  \eqref{cons4} can be re-expressed by
 \begin{align}
&{\cal I}_{xy}\left(\bf V\right)=M\left(\sum\nolimits_{b = 1}^B {( {{\rm{Tr}}( {{\bf{\hat G}}_{bxy}^{\rm{I}}{\bf{V}}_b^{\rm{R}}} ) + {{\bar \kappa }_{bxy}}{\rm{Tr}}( {{\bf{V}}_b^{\rm{R}}} )} )}\right. \nonumber\\
 &\left.+ \sum\nolimits_{k = 1}^K {{\rm{Tr}}( {{\bf{\tilde G}}_{xy}^{\rm{I}}{\bf{V}}_k^{\rm{C}}} )}  \right)\le I_{xy}^{\max } ,\forall (x,y) \in {{\cal S}_o}, \tag{\rm C18} \label{cons18}
\end{align}
where ${\bf{\hat G}}_{xy}^{\rm{I}}={{\bf{g}}_{bxy}^{\rm{L}}}({{\bf{g}}_{bxy}^{\rm{L}}})^{\rm H}$ and ${\bf{\tilde G}}_{xy}^{\rm{I}} = \left( {{\bf{G}}_{xy}^{\rm{I}} + \sum\nolimits_{b = 1}^B {\left( {{{\bar \kappa }_{bxy}}{{\bf{B}}_b}} \right)} } \right)$ with denoting   ${\bf{ G}}_{xy}^{\rm{I}} = {\bf{g}}_{xy}{ {\bf{g}}_{xy} ^{\rm{H}}}$ and ${\bf{g}}_{xy} = [ {{{( {{\bf{g}}_{1xy}^{\rm{L}}} )}^{\rm{T}}},\ldots,{{( {{\bf{g}}_{Bxy}^{\rm{L}}} )}^{\rm{T}}}} ]^{\rm{T}}$.

Then, with ${U_{\rm{C}}}\left( {{{\bf{z}}^{\rm{C}}}} \right)$ defined in \eqref{eq330} and $z^{\rm R}$ defined in \eqref{cons370}, the objective function of problem \ref{eqp1} is rewritten by
\begin{align}
\!\!\!\!\ln U \! =\! {\alpha _{\rm{C}}}\ln {U_{\rm{C}}}\left( {{{\bf{z}}^{\rm{C}}}} \right) \!+\! {\alpha_{\rm{R}}}\ln {z^{\rm{R}}} \!-\! { {{\alpha _{\rm{A}}}}}\ln  { \sum\nolimits_{b = 1}^B {{a_b}{o_b}} }.\label{eq38}
\end{align}

Then, based on \eqref{cons7}, and defining the auxiliary constants ${o_{\min }} =\ln \left( {{{\min }_{1 \le b \le B}}\;{o_b}} \right)$ and $
 {o_{\max }}  =\ln ( {\sum\nolimits_{b = 1}^B {{o_b}} } ) $, we have
 \begin{align}
 {o_{\min }} \le \ln \left( {\sum\nolimits_{b = 1}^B {{a_b}{o_b}} } \right) \le  {o_{\max }}.
  \end{align}

By introducing an auxiliary optimization variable $z^{\rm{A}}$, equation \eqref{eq38} can be reformulated as:
\begin{align}
\ln U = &\mathop {\sup }\limits_{{z^{\rm{A}}}} \left\{ {\cal U}\left( {\bf{z}} \right) \buildrel \Delta \over = {{\alpha_{\rm{C}}}\ln {U_{\rm{C}}}\left( {{{\bf{z}}^{\rm{C}}}} \right)\; + {\alpha_{\rm{R}}}\ln {z^{\rm{R}}}} \right.\nonumber\\
&\qquad\qquad\qquad\qquad\qquad\left. { + {\alpha_{\rm{A}}}\left( {{z^{\rm{A}}} - {o_{\max }}} \right)  } \right\}  \label{eq390}\\
&\quad   {\rm{s}}{\rm{.t.}}\;  0 \le {z^{\rm{A}}} \le {o_{\max }}- o_{\min} ,  \tag{\rm C19} \label{cons19}\\
&\quad  \quad\;\;  \sum\nolimits_{b = 1}^B {{a_b}{o_b}}  \le {e^{{o_{\max }} - {z^{\rm{A}}} }} \buildrel \Delta \over = \Upsilon \left( {\bf{z}} \right),  \tag{\rm C20} \label{cons20}
 \end{align}
 where ${\bf{z}} = [{{\bf{z}}^{\rm{C}}},{z^{\rm{R}}},{z^{\rm{A}}}]\in{\mathbb C}^{1\times(K+2)}$.

Finally,  with \eqref{eq330}, \eqref{cons370}, and ${\cal U}\left({\bf{z }}\right)$ defined in \eqref{eq390}, problem \ref{eqp1} can be equivalently rewritten as:
\begin{align}
\mathop {\max }\limits_{{\bf{V}},{\bf{z }},{\bf{a}},{\bm \epsilon}}\quad&\; {\cal U}\left({\bf{z }}\right)\tag{\bf P1}\label{eqp2}\\
{\rm{s}}{\rm{.t}}{\rm{.}}\quad &\eqref{cons6}-\eqref{cons14}, \eqref{cons17}-\eqref{cons20}. \nonumber
\end{align}

It can be observed that the objective function of  \ref{eqp2} is monotonically increasing with $\bf z$.
However,  \ref{eqp2} remains non-convex due to non-convex constraints
 \eqref{cons8}, \eqref{cons20},  and \eqref{cons10}, and the integer constraint \eqref{cons6}.
 Therefore, it still belongs to the MINLP and is challenging to solve.
Note that the conventional Benders' decomposition method  \cite{bi2019joint, chen2020joint} can be applied to solve MINLP problems.
However, it necessitates that the remaining continuous optimization problem is convex when the integer variables are fixed.
Unfortunately,   \ref{eqp2} does not satisfy this condition, which prompts the need for developing new algorithms to find optimal solutions.

\section{Optimal Solution: MO-BRB Optimization}

In this section, we propose the following monotonic optimization
embedded branch-and-bound (MO-BRB) algorithm to find the optimal solution of problem \ref{eqp2}.

\subsection{MO-BRB Algorithm Development}

To solve problem \ref{eqp2}, we need to address the non-convexity that arises from constraints \eqref{cons8} and \eqref{cons20}, the rank constraint \eqref{cons10}, and the integer constraint \eqref{cons6}. Moreover, we have the following strategies during the algorithm development:
 \begin{itemize}
 \item{\bf Strategy 1}: The BRB algorithm can address integer programming problems by iteratively branching them into subproblems with a mix of fixed integer solutions and relaxed continuous convex sets. For example, in each branch, the integer constraint \eqref{cons6} can be relaxed as the following convex constraint, i.e., ${\bf a}\in{\cal A} \buildrel \Delta \over = {{\cal D}_1} \times {{\cal D}_2}\times\ldots \times {{\cal D}_B}$, where
     ${{\cal D}_b} $ is the $b$-th sub-domain regarding to ${a_b}$  and is selected from one of the following convex sets: ${{\cal D}_b} = \left\{ {{a_b}\left| {  {a_b} = 0} \right.} \right\}$, ${{\cal D}_b} = \left\{ {{a_b}\left| {  {a_b} = 1} \right.} \right\}$, or ${{\cal D}_b} = \left\{ {{a_b}\left| {0 \le {a_b} \le 1} \right.} \right\}$.

 \item{\bf Strategy 2}: By fixing $\bf{z}$ and relaxing the integer constraint \eqref{cons6} to a  convex set, constraints \eqref{cons8} and \eqref{cons20} become convex in $\bf{W}$.
This simplifies the remaining problem to optimize the continuous variables $\bf{a}$ and $\bf{W}$, which includes the only non-convex rank constraint \eqref{cons10}.
Fortunately, the optimal solution to this problem can be found without considering \eqref{cons10}, since
Section \ref{secAlgorithm} will show that the optimal solution can automatically satisfy \eqref{cons10}. Hence, \eqref{cons10} can be disregarded during the optimization process.

 \item{\bf Strategy 3}: The objective function is monotonically increasing in $\bf{z}$. Based on this condition,
             if the optimal solutions for the other variables can be found with low complexity when $\bf{z}$ is fixed, the MO optimization method can be used to determine the optimal ${\bf z}$.
\end{itemize}

Therefore, the two-stage MO-BRB algorithm is employed to find the optimal solution of \ref{eqp2}, wherein the outer stage utilizes the BRB to address the integer optimization, while the inner stage applies the MO to handle the continuous optimization.

Specifically, with {\bf Strategy 1}, we apply the BRB  to iteratively branch \ref{eqp2} into subproblems with a mix of fixed integer solutions and continuous convex sets,   i.e., ${\bf{a}}\in {\cal A}$. For example, in each branch, we solve the following subproblem without integer constraints, i.e.,
\begin{align}
\!\!{\cal Q}_{ \cal A} =  \;\mathop {\max }\limits_{ {\bf{z}} \in {\cal Z} }&\; {\cal U}\left( {{\bf{z}} } \right)\tag{\bf P2} \label{eqp3}\\
{\rm{s}}{\rm{.t}}{\rm{.}} & \;{\cal Z}  = \left\{ {{\bf{z}}\left| \eqref{cons7}-\eqref{cons14}, \eqref{cons17}-\eqref{cons20}  \right.}, {\bf{a}}\in {\cal A}\right\}, \nonumber
  \end{align}
whose optimal solution can be obtained by applying the MO method  \cite{bjornson2013optimal} due to {\bf Strategies 2-3}.
Then, the BRB algorithm can systematically reduce the search space through the imposition of bounds and strategically bound feasible solutions, leading to efficient convergence towards an optimal solution. In addition, we summarize the MO-BRB in Algorithm \ref{algorithmBRB}.

In the following, we introduce the MO method to solve the subproblem \ref{eqp3}  that appears in the inner-stage of each branch of BRB (step 12 of Algorithm \ref{algorithmBRB}).

\subsection{Preliminaries of Monotonic Optimization (MO) }\label{MOA}
This part introduces the preliminaries of MO \cite{bjornson2013optimal,chen2025otfs}.
\begin{itemize}
\item {\bf Box}:
Given any two $N$-dimensional vectors ${\bf{z}}^{\rm min}\in{\mathbb R}_{+}^{N}$ and ${\bf z}\in{\mathbb R}_{+}^{N}$, whose elements  satisfy ${\left[ {{{\bf{z}}^{{\rm{min}}}}} \right]_n} \le {\left[ {\bf{z}} \right]_n}$ for $1\le n \le N$,
 the hyper rectangle $\left[ {{\bf{z}}^{\rm min},{\bf{z}}} \right] = \left\{ {{\bf{x}}\left| {{\bf{z}}^{\rm min} \le {\bf{x}} \le {\bf{z}}} \right.} \right\}$ is defined as the box with the lower vertex ${\bf{z}}^{\rm min}\in{\mathbb R}_{+}^{N}$ and the upper vertex ${\bf z}\in{\mathbb R}_{+}^{N}$.

\item {\bf Normal}: An infinite set ${\cal Z} \subset {\mathbb R}_{+}^{N} $ is normal if given any element ${\bf z}\in{\cal Z}$, we have
the box $\left[ {{\bf{z}}^{\rm min},{\bf{z}}} \right]\subset {\cal Z}$.
\item {\bf Polyblock}: Denoting a finite set ${\cal V}\subset{\mathbb R}_{+}^{N}$, the union of all boxes $\left[ {{\bf{z}}^{\rm min},{\bf{z}}} \right] $, ${\bf z}\in{\cal V}$
is named  the  polyblock ${\cal B}({\bf{z}}^{\rm min},\cal V)$ with the lower vertex ${\bf{z}}^{\rm min}$ and the upper vertex set $\cal V$.
\item {\bf Projection}: For any non-empty normal set ${\cal Z} \subset {\mathbb R}_{+}^{N} $ and any vertex ${\bf z}$, the projection of $\bf z$ onto the boundary of $\cal Z$ is denoted by ${\rm{Proj}}\left( {\bf{z}} \right) = {{\bf{z}}^{{\rm{min}}}} + \delta^\star \left( {{\bf{z}} - {{\bf{z}}^{{\rm{min}}}}} \right)$, where ${\delta ^\star} = \max \left\{ {\delta \left| {{{\bf{z}}^{{\rm{min}}}} + \delta \left( {{\bf{z}} - {{\bf{z}}^{{\rm{min}}}}} \right) \in {\cal Z},0 \le \delta  \le 1} \right.} \right\}$.
\item {\bf MO problem}: A problem belongs to the MO family if it satisfies the following forms, i.e.,
   \begin{align}
\mathop {\max }\limits_{\bf{z}}\;\; {\cal W}\left( {\bf{z}} \right), \quad {
\rm s.t.}\; {\bf{z}} \in {\cal Z}, \label{MO}
\end{align}
where $ {\cal W}\left( {\bf{z}} \right)$  is an increasing function of the vertex $\bf z$ on ${\mathbb R}_{+}^{N}$ and $ {\cal Z}$ is a non-empty normal closed set.
\end{itemize}

\begin{algorithm}[t]
\caption{MO-BRB optimization for   \ref{eqp2}}\label{algorithmBRB}
 \renewcommand{\baselinestretch}{1}
 {
{\bf  RELAXATION: (step 1-step 4)}\\
	 Initialize ${\cal A} \buildrel \Delta \over = {{\cal D}_1} \times {{\cal D}_2}\times\ldots \times {{\cal D}_B}$  using  convex sub-domains, where ${{\cal D}_b} = \left\{ {{a_b}\left| {0 \le {a_b} \le 1} \right.} \right\}$\;
Initialize the set of relaxed domains by ${\cal A}_{\rm all}=\left\{{\cal A} \right\}$ \;

Initialize the current optimal branching domain ${{\cal A}^\dag}={\cal A}$, the lower bound ${\cal Q}^{\rm lb}=-\infty$, the upper bound ${\cal Q}^{\rm ub}= \infty$,  and the maximum number of iterations $\ell_{\rm max} = 2^B$ \; 

	 \For{$\ell\leftarrow 1$ \KwTo $\ell_{\rm max}$}{
{\bf  BRANCHING: (step 6-step 12)}\\

\lIf{sub-domain $b$ of ${\cal A}^\dag$ is an integer domain for any arbitrary $b$, i.e.,  ${\cal D}_b^\dag \subset {\left\{ { \left\{0\right\}, \left\{1\right\}} \right\}}, \forall b$}
 {\\Terminate and return the optimal ${\bf a}^\star$}
\Else  {Select an arbitrary non-integer (continuous) sub-domain, i.e.,  ${\cal D}_b^\dag \not\subset {\left\{ { \left\{0\right\}, \left\{1\right\}} \right\}}$ and define the following two new branches:
 \begin{align}
{\cal A}_0^\dag  = {\cal D}_{1}^ \dag  \times {\cal D}_{2}^\dag\ldots  \times \left\{ {{a_b} = 0} \right\} \ldots  \times {\cal D}_{B}^\dag,\nonumber\\
{\cal A}_1^\dag = {\cal D}_{1}^\dag  \times {\cal D}_{2}^ \dag \ldots \times \left\{ {{a_b} = 1} \right\} \ldots  \times {\cal D}_{B}^ \dag.\nonumber
\end{align}

 \For{$i\leftarrow 0$ \KwTo $1$}{
 {{\bf Inner-stage Optimization:} Given ${\cal Q}^{\rm lb}$ and ${\cal A}={\cal A}_i^\dag $, solve the MO problem \ref{eqp3} by Algorithm \ref{algorithmMO}, and denote its solution and objective function value by ${\bf a}_{{\cal A}_i^\dag}$ and ${\cal Q}_ {{\cal A}_i^\dag} $ \;
}}}
{\bf  BOUNDING: (step 13-step 22)}\\
Remove ${\cal A}^\dag$ from ${{\cal A}_{\rm all}}$, i.e.,  ${{\cal A}_{\rm all}}\leftarrow  {{{\cal A}_{\rm all}}\backslash {{\cal A}^ \dag }} $\;
 \For{$i\leftarrow 0$ \KwTo $1$}
 {
 \If  {  ${\cal Q}_ {{\cal A}_i^\dag}  >{\cal Q}^{\rm lb}$  }{
  \If { ${\bf a}_{{\cal A}_i^\dag}$ are integer solutions }{
 Update ${\cal Q}^{\rm lb} ={\cal Q}_{{\cal A}_i^\dag}$, ${\bf a}^\star={\bf a}_{{\cal A}_i^\dag}$, and $
  {\cal A}_i^\dag \leftarrow \left\{ {\left[ {{\bf a}^\star} \right]_1} \right\}  \times \left\{ { \left[ {{\bf a}^\star} \right]_2} \right\}  \ldots \times \left\{ {\left[ {{\bf a}^\star} \right]_B} \right\}$\tcp*[f]{Because ${\bf a}_{{\cal A}_i^\dag}$ is superior to any sub-branches of the old ${\cal A}_i^\dag$ }
}
   ${{\cal A}_{\rm all}}\leftarrow {{\cal A}_{\rm all}}\bigcup \left\{{\cal A}_i^\dag \right\}$\;}
}
  \For {$\forall {\cal A} \in {\cal A}_{\rm all}$}
   { \If {${\cal Q}_ {{\cal A}}  <{\cal Q}^{\rm lb}$}
   {Delete this suboptimal branch: ${{\cal A}_{\rm all}}\leftarrow {{\cal A}_{\rm all}}\backslash \left\{{\cal A} \right\}$ }}
Update  the upper bound by ${\cal Q}^{\rm ub} =    Q_{{\cal A}^\dag}$, where ${\cal A}^\dag$ is the current optimal branch, i.e.,
$
 {\cal A}^\dag = \mathop {\arg \max }\nolimits_{{\cal A} \in {\cal A}_{\rm all}} {\cal Q}_{ \cal A}$
}
}
\end{algorithm}

\subsection{Optimal Solution to the MO Problem \ref{eqp3}}
It is obvious that problem \ref{eqp3} belongs to the MO family when ignoring \eqref{cons10} as explained in {\bf Strategies 2-3}, whose optimal solutions can be obtained by applying the MO techniques, i.e., the outer polyblock algorithm  \cite{bjornson2013optimal,chen2025otfs}.

\begin{figure*}[ht]
\centering
\includegraphics[width = 0.95\textwidth]{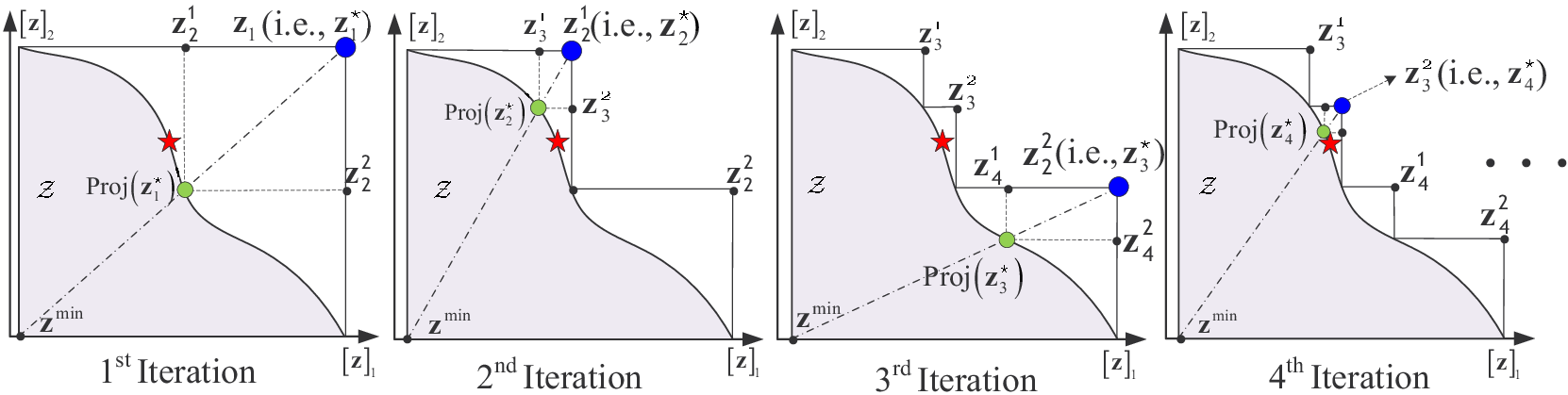}
\caption{Illustration of the outer polyblock algorithm when  $\bf z$ is a two-dimensional vector ($N=2$): the red star represents the optimal global solution,  the blue circle denotes the optimal vertex in each iteration, and the green circle indicates the projected point on the boundary.} \label{fig2}
\end{figure*}

Before introducing algorithm details, we explain the principles behind how the outer polyblock algorithm solves MO problems.
Firstly, since the objective function is monotonically increasing with  $\bf z$,  the optimal solution of  $\bf z$ must lie on the boundary of the feasible region $\cal Z$. Therefore, as shown in Fig \ref{fig2}, we can iteratively utilize multiple polyblocks to approximate the feasible boundary  $\cal Z$.
Then, in each iteration, we shrink the polyblock space by branching polyblocks and discarding those that fall outside the feasible region, resulting in a closer distance between the boundary and the outer of the polyblock space, allowing us to find the optimal solution. Below, we outline the main steps of the outer polyblock algorithm.

\subsubsection{\bf Initialization}
In problem \ref{eqp3}, we have $N=K+2$ because we define ${\bf{z}} = [{{\bf{z}}^{\rm{C}}},{z^{\rm{R}}},{z^{\rm{A}}}]\in{\mathbb C}^{1\times(K+2)}$.
By analyzing \eqref{cons8}, \eqref{cons9}, \eqref{cons12}, and \eqref{cons20},
we know the feasible set $\cal Z$ is the subset of box  $\left[ {{{\bf{z}}^{{\rm{min}}}},{{\bf{z}}^{{\rm{max}}}}} \right]$, i.e., $\cal Z\subset\left[ {{{\bf{z}}^{{\rm{min}}}},{{\bf{z}}^{{\rm{max}}}}} \right]$, where the $n$-th elements of ${\bf z}^{\rm max}$ and  ${\bf z}^{\rm min}$  are, respectively, given by
 \begin{align}
\!\!\!{\left[ {{{\bf{z}}^{{\rm{max}}}}} \right]_n }& \!=\! \left\{ {\begin{array}{*{20}{l}}
\!\!{\ln \left( {1\! + \!\frac{{\mathop {\max }\limits_b P_b^{\max }}}{{{\sigma _n}}}{\rm{Tr}}\left( {{\bf{G}}_n^{\rm{C}}} \right)} \right),   1 \le n  \le K,}\\
{\sum\nolimits_{q = 1}^Q {w_q^{\rm{R}},    n  = K + 1,} }\\
{1,   n  = K + 2,}
\end{array}} \right.\\
\!\!\!{\left[ {{{\bf{z}}^{{\rm{min}}}}} \right]_n }& \!=\! \left\{ {\begin{array}{*{20}{l}}
{C_n^{\min }\ln 2,  1 \le n  \le K,}\\
{0,   n  = K + 1,}\\
{0,   n  = K + 2.}
\end{array}} \right.
 \end{align}
Then, we establish an initial polyblock ${\cal B}({\bf z}^{\min}, {\cal V}_1)$ that contains the whole feasible set $\cal Z$ with the  vertex set ${\cal V}_1=\left\{{{\bf z}^{\rm max}}\right\}$.

\subsubsection{\bf  Vertex Finding}
In the $i$-th iteration, we find  the optimal polyblock vertex in ${\cal V}_{i}$  that maximizes the objective function:
\begin{align}
{\bf z}_{i}^\star=\arg \mathop {\max }\limits_{{\bf{z}} \in {\cal V}_{i}}\;\; {\cal U}\left( {\bf{z}} \right).
 \end{align}

\subsubsection{\bf  Projection}
We project  ${\bf z}_{i}^\star$ on to the feasible boundary $\cal Z$, i.e., ${\rm{Proj}}\left( {{\bf{z}}_{i }^\star} \right)$ as defined in the preliminaries of MO.

\subsubsection{\bf Polyblock Reduction and Bounding}
With ${\rm{Proj}}\left( {{\bf{z}}_{i }^\star} \right)$, we can construct a smaller polyblock ${\cal B}({\bf z}^{\min}, {\cal V}_{i+1})$, where the upper vertex set ${\cal V}_{i+1}$  is obtained by deleting ${\bf z}_{i}^\star$ from  set ${\cal V}_{i}$ and adding $(N=K+2)$ new vertices  into this set, i.e.,
 \begin{align}
 {\cal V}_{i+1}=\left({\cal V}_{i}\backslash \left\{{{\bf z}_{i }^\star}\right\}\right) \cup \left\{ {{\bf z}_{i+1}^1,{\bf z}_{i+1}^2,\ldots,{\bf z}_{i+1}^{K+2}} \right\},\label{eqaddv}
\end{align}
where  ${\bf{z}}_{i+1}^n\le{\bf{z}}_{i }^\star$ and
 \begin{align}
{\bf{z}}_{i+1}^n = {\bf{z}}_{i }^\star - \left( {{{\left[ {{\bf{z}}_{i }^\star - {\rm{Proj}}\left( {{\bf{z}}_{i }^\star} \right)} \right]}_{n}}} \right){{\bf{e}}_n },1 \le n  \le K + 2.\label{eqproj1}
\end{align}
Here, ${\bf e}_n\in{\mathbb C}^{(K+2)\times1}$ is a unit vector with a value of one in the $n$-th element and zero in all other elements.
Besides, if the  $n$-th element of ${\bf{z}}_{i}^\star $ is  close to ${\left[{\bf z}^{\rm min} \right]}_{n}$, there is no need to include  ${\bf{z}}_{i+1}^n $ into ${\cal V}_{i+1}$ for a better convergence.

Note that as the iteration continues, the feasible set $\cal Z$ that includes the potential optimal solution is always contained in the new polyblock with the vertex set $ {\cal V}_i$, and we have
${\cal B}({\bf z}^{\min}, {\cal V}_{1})\supset{\cal B}({\bf z}^{\min}, {\cal V}_{2}) \ldots\supset{\cal B}({\bf z}^{\min}, {\cal V}_{i})\supset{\cal Z}$.
Furthermore, the upper bound achieved in each iteration, i.e., $U^{\rm ub}\leftarrow{\cal U}\left(
{\bf z}_{i}^\star \right)$ will iteratively decrease due to the smaller polyblock.
Also, if ${\cal U}\left(
{\rm{Proj}}\left( {{\bf{z}}_{i }^ \star } \right) \right)$ is larger than the current lower bound $U^{\rm lb}$, we can update it by $U^{\rm lb}\leftarrow {\cal U}({\rm Proj}({\bf z}_i^\star))$. Finally,  the iteration will terminate if $U^{\rm ub}\le(1+\tilde \rho  )U^{\rm lb}$, where the positive parameter $\tilde \rho  $ represents the error tolerance that determines the desired level of accuracy for the approximation.
Finally, we show  how to find ${{\rm{Pro}}{{\rm{j}} }\left( {{\bf{z}}_{i }^*} \right)}$ in the next subsection.

\subsection{Projection in Polyblock Algorithm }\label{secAlgorithm}
From the definition of projection in Section~\ref{MOA}, we know
\begin{align}
{\rm{Proj}}\left( {{\bf{z}}_{i }^ \star } \right) = {{\bf{z}}^{\min }} + {\delta ^ \star }\left( {{\bf{z}}_{i}^ \star  - {{\bf{z}}^{\min }}} \right),
\label{eqproj0}
 \end{align}
where ${\delta ^\star}$ represents the maximum value such that ${\rm{Proj}}\left( {{\bf{z}}_{i - 1}^ \star } \right)$ remains a feasible solution for problem \ref{eqp3}, i.e.,
\begin{align}
{\delta ^\star} &= \max \left\{ {\delta \left| {{{\bf{z}}^{{\rm{min}}}} + \delta \left( {{\bf{z}} - {{\bf{z}}^{{\rm{min}}}}} \right) \in {\cal Z},0 \le \delta  \le 1} \right.} \right\}\nonumber\\
 &= \mathop {\arg \max }\limits_{0\le\delta\le1} \;\delta \tag{\bf P3}\label{eqP4}\\
&\quad{\rm{s}}{\rm{.t}}{\rm{.}}\; \left( {\exp \left( {{{\left[ {{{{\bf{\hat z}}}_\delta }} \right]}_k}} \right) - 1} \right){{\cal H}_k}\left( {\bf{V}} \right) \le {{\cal F}_k}\left( {\bf{V}} \right),\forall k, \tag{\rm C21} \label{cons21}\\
&\qquad\;\; {\left[ {{{{\bf{\hat z}}}_\delta }} \right]_{K + 1}} \le \sum\nolimits_{q = 1}^Q {w_q^{\rm{R}}\epsilon_q^{\rm{R}}},  \tag{\rm C22} \label{cons22}\\
&\qquad \;\;\sum\nolimits_{b = 1}^B {{a_b}{o_b}}  \le {e^{{o_{\max }}\left( {1 - {{\left[ {{{{\bf{\hat z}}}_\delta }} \right]}_{K + 2}}} \right)}},  \tag{\rm C23} \label{cons23} \\
&\qquad\;\;  \eqref{cons7}, \eqref{cons10}-\eqref{cons13},\eqref{cons17}-\eqref{cons19}, {\bf a}\in{\cal A}, \nonumber
\end{align}
where  ${\bf \hat z}_\delta={{\bf{z}}^{\min }} + {\delta   }\left( {{\bf{z}}_{i }^ \star  - {{\bf{z}}^{\min }}} \right)$. Here, constraints \eqref{cons8}-\eqref{cons9} are replaced by \eqref{cons21}; constraints \eqref{cons14} and  \eqref{cons20}   are replaced by  \eqref{cons22} and  \eqref{cons23}, respectively.

Next, since the objective function of \ref{eqP4} is monotonic, continuous, and bounded, we can use the binary search to find the maximum $\delta$, as shown in Algorithm~\ref{algorithmProj3}.
 Specifically,  upon denoting the value of $\delta$ in the $j$-th iteration of the binary search by $\delta_j$,  we need to solve the following feasibility checking problem to decide whether to increase or decrease  $\delta$, i.e.,
\begin{align}
\mathop {\max }\limits_{{\bf{a}},{\bf{V}},{\bm \epsilon}} &\;1\tag{\bf P4} \label{eqp5} \\
{\rm{s}}{\rm{.t}}{\rm{.}}\;& \delta=\delta_j\;  {\rm and} \;\eqref{cons21}-\eqref{cons23}.\\
&\eqref{cons7}, \eqref{cons10}-\eqref{cons13},\eqref{cons17}-\eqref{cons19}, {\bf a}\in{\cal A}. \nonumber
\end{align}
However, this problem is still non-convex due to \eqref{cons10}. To handle such non-convexity,  we usually apply the semidefinite relaxation (SDR) technique \cite{li2021many} to ignore this constraint and apply the randomization-based method \cite{long2019symbiotic} to obtain the approximated rank one solution.
In this paper, we propose the following theorem to show that if the relaxed problem of \ref{eqp5} is feasible, there always exist optimal solutions to  \ref{eqp5} that satisfy \eqref{cons10}. As a result,  \eqref{cons10} can be disregarded.

\begin{theorem}\label{theoremrank1}
If the relaxed formulation of   \ref{eqp5} by ignoring constraint \eqref{cons10} is feasible, there always exists a globally optimal solution with a rank of one or less. This solution can therefore be considered the optimal solution to   \ref{eqp5}.
\end{theorem}
\begin{IEEEproof}
Please refer to Appendix \ref{appendix2}.
\end{IEEEproof}
Since the relaxed problem of \ref{eqp5} ignoring  \eqref{cons10} is convex, it can be solved by Matlab toolbox CVX within polynomial complexity \cite{chen2019optimal}. Then, we can obtain the optimal solution of  \ref{eqp5} by applying the method mentioned in Appendix \ref{appendix2} that satisfies rank constraint \eqref{cons10}.
Finally, we provide an overview of the polyblock algorithm for solving the MO problem in Algorithm \ref{algorithmMO}.
Additionally, the binary search algorithm used to solve the projection problem in \ref{eqP4} during the iterations of the polyblock algorithm is summarized in Algorithm \ref{algorithmProj3}.

\begin{algorithm}[t]
\caption{Outer polyblock algorithm for   \ref{eqp3}}\label{algorithmMO}
 \renewcommand{\baselinestretch}{0.95}
 {
\If{ {\rm problem} \ref{eqP4} with setting ${\bf z}={\bf z}^{\rm min}$is  feasible}{
{\bf INITIALIZATION: } \\
Initialize iteration index $i=1$, polyblock ${\cal B}({\bf z}^{\min}, {\cal V}_1)$ with vertex set ${\cal V}_1=\left\{{{\bf z}^{\rm max}}\right\}$ \;
Set the lower bound, upper bound, error tolerance by $U^{\rm lb}={\cal Q}^{\rm lb}$, $U^{\rm ub}=U( {{\bf z}^{\rm max}})$, $\tilde \rho_1=0.05$, respectively,  where ${\cal Q}^{\rm lb}$ is obtained in steps 4 or 18 of Algorithm 1\;
\While{ $U^{\rm ub}>(1+\tilde \rho  )U^{\rm lb} $ }{

{\bf VERTEX FINDING:} \\
Find  the current optimal vertex in ${\cal V}_{i}$ that maximizes the objective function:
\begin{align}
{\bf z}_{i}^\star= \arg \mathop {\max }\nolimits_{{\bf{z}} \in {\cal V}_{i}} {\cal U} \left( {\bf{z}} \right) \nonumber
\end{align}

{\bf PROJECTION: } \\

Given ${\bf z} = {\bf z}_{i}^\star$, solve projection problem \ref{eqP4} and obtain ${\rm Proj}({\bf z}_{i}^\star)$ and $\delta^\star$ by using Algorithm \ref{algorithmProj3}\;

{\bf  POLYBLOCK REDUCTION:} \\ 

Calculate $N=K+2$ new vertices ${\bf{z}}_{i+1}^n$ by  \eqref{eqproj1} and update the new vertex set ${\cal V}_{i+1}$ by \eqref{eqaddv}\;
Construct a smaller polyblock ${\cal B}({\bf z}^{\min}, {\cal V}_{i+1})$\;

{\bf  BOUNDING: } \\ 
Update the lower bound with the feasible inner point: ${{ U}^{{\rm{lb}}}} \leftarrow \max \left( {{\cal U}({\rm{Proj}}({\bf{z}}_i^ \star )),{U^{{\rm{lb}}}}} \right)$

Update  the upper bound by ${ U}^{\rm ub} \leftarrow {\cal U}({\bf z}_{i}^\star)$\;


$i=i+1$\;
}
}}
\end{algorithm}

\begin{algorithm} [t]
\caption{Bisection algorithm for   \ref{eqP4}}\label{algorithmProj3}
\renewcommand{\baselinestretch}{0.95}
{
Initialize iteration index $j=0$\;
Set $\delta_{\rm max}=0$, $\delta_{\rm max}=1$, and   $\tilde \rho_2\ll1$ \;
\While{ $\delta_{\rm max}-\delta_{\rm min}\ge\tilde \rho_2$ }{
    $j=j+1, \delta_j=\frac{1}{2}\left(\delta_{\rm max}+\delta_{\rm min}\right)$\;
    Check the feasibility of the relaxed problem of \ref{eqp5} by CVX \;
    \lIf {Feasible}{$\delta_{\rm min}=\delta_j$}
    \lElse {${\delta_{\rm max}=\delta_j}$}
    }
}
\end{algorithm}

\subsection{Performance Analysis}

\subsubsection{Algorithm Optimality and Convergence}
As mentioned earlier, the optimal solution of  \ref{eqp2} can be obtained by jointly using the BRB  in the outer stage for integer optimization and the MO  in the inner stage for continuous optimization. The detailed reason is provided as follows.

In the outer-stage BRB, problem \ref{eqp2} is iteratively divided into subproblems by branching on the unknown integer variables.
Each subproblem consists of a mix of known fixed integer solutions and unknown relaxed continuous convex sets.
Assuming the optimal solution of any such subproblem can be determined, the initial upper bound of problem \ref{eqp2} corresponds to the case where all integer variables are relaxed, while the initial lower bound is given by any subproblem with fixed integer variables.
During each iteration, subproblems with bounds below the current lower bound are pruned, the lower bound is updated if a better integer solution is found, and subproblems with relaxed integer variables are further branched to refine the upper bound.
Finally, the upper bound decreases toward the increasing lower bound associated with the integer solution, thereby ensuring both the convergence and optimality of the BRB algorithm.

In the inner-stage MO, as illustrated in Fig. \ref{fig2}, the branched subproblem is solved by iteratively approximating the feasible boundary  $\cal Z$ using a polyblock.
At each iteration, the polyblock space is shrunk through branching and discarding the regions outside the feasible set, thereby reducing the difference between the feasible boundary and the outer edge of the polyblock.
Meanwhile, the lower bound, determined by the projection of a vertex, and the upper bound, obtained at the vertex of the outer polyblock, gradually converge.
This process guarantees the convergence and optimality of the MO method.

Therefore, we can conclude that the convergence and optimality of the MO-BRB algorithm are guaranteed.

\subsubsection{Algorithm Complexity}The complexity of the MO-BRB algorithm is jointly determined by each algorithm.
The BRB algorithm iteratively branches problem \ref{eqp2} into subproblems and eventually reaches a solution with integer variables, resulting in a worst-case iteration number of $2^{B}$. For the MO, we iteratively tighten the outer polyblock, with each iteration involving the application of a bisection algorithm and solving problem \ref{eqp5}.  Specifically, from \cite{nesterov1994interior}, the complexity of solving a convex problem can be generally given by ${\cal O}\left({\cal N}_C^{1.5}{\cal N}_D^{2}\right)$, where ${\cal N}_C=2\left( {K + B} \right) + 8Q+7 + \left| {{{\cal S}_o}} \right|$ and ${\cal N}_D=B+3Q+\left( {B + K{B^2}} \right)N_{\rm tx}^2$
represent the number of constraints and the dimension of optimization variables of problem \ref{eqp5}, respectively.
Finally, upon denoting the number of iterations required for the outer polyblock and bisection algorithms to converge as ${\cal N}_{\rm P}$ and ${\cal N}_{\rm B}$, respectively,  the complexity of the MO-BRB can be generally given by  ${\cal O}\left(2^B{\cal N}_{\rm P}{\cal N}_{\rm B}{\cal N}_C^{1.5}{\cal N}_D^{2}\right)$.

\section{Low-Complexity Solution: SCA Optimization}
In this section, we develop a low-complexity algorithm to efficiently solve problem \eqref{eqp2} with near-optimal performance.

\subsection{ SCA Algorithm Development}
To address this MINLP problem, we first relax integer constraints using a penalized method and subsequently transform the optimization problem into a tractable formulation, which can be solved iteratively by the SCA algorithm \cite{shan2024resource}.
Specifically, during each iteration of  SCA, the non-convex terms are approximated as affine terms using a Taylor series expansion.
This approximation converts the non-convex problem into a convex one, which can then be efficiently solved using convex optimization toolboxes.

To begin with, we replace the integer constraint \eqref{cons6} by  using the following equivalent constraints:
\begin{align}
&  {{a_b} - a_b^2}   \le 0, \forall b,\tag{\rm C24} \label{cons24} \\
&0 \le {a_b} \le 1,\forall b.    \tag{\rm C25} \label{cons25}
\end{align}

With constraints \eqref{cons24} and \eqref{cons25}, we only have the continuous optimization variables now.
Besides, considering the  non-convex constraint \eqref{cons24}, we introduce a penalty factor $ {\hat \eta } $ to reformulate problem \ref{eqp2} as the following  problem:
\begin{align}
\mathop {\max }\limits_{{\bf{V}},{\bf{z }},{\bf{a}}}\;&\;  {\cal U}\left({\bf{z }}\right)-{\hat \eta }\left({\cal G}\left( {\bf{a}} \right) - \hat {\cal G}\left( {\bf{a}} \right)\right)
\tag{\bf P6},\label{eqP6}\\
{\rm{s}}{\rm{.t}}{\rm{.}}\; &\eqref{cons7}-\eqref{cons14}, \eqref{cons17}-\eqref{cons20},   \nonumber
  \end{align}
where the constant penalty factor $ {\hat \eta } \gg 1 $ can penalize the objective function when $a_b$ is not equal to 0 or 1, and
 \begin{align}
{\cal G}\left( {\bf{a}} \right) = \sum\nolimits_{b = 1}^B {{a_b}} ,\hat {\cal G}\left( {\bf{a}} \right) = \sum\nolimits_{b = 1}^B {a_b^2}.
\end{align}

However,  \ref{eqP6} is still non-convex due to  constraints \eqref{cons8}, \eqref{cons20}, $\hat {\cal G}\left( {\bf{a}} \right)$, and the rank constraint \eqref{cons10}.
Given that problem \ref{eqP6} is a maximization problem, an approximate problem can be formulated by adopting a concave lower-bound objective function and imposing tighter convex constraints compared to problem \ref{eqP6}.

To do so, we denote the optimized solution in the $i$-th iteration of SCA by  ${\bf a}_i$ and
${\bf{z}}_i = [z _{1,i}^{\rm{C}},\ldots,z _{K,i}^{\rm{C}} ,{z_i^{\rm{R}}},{z_i^{\rm{A}}}]$.
Then, in the $(i+1)$-th iteration, we utilize the first-order Taylor series expansion to approximate them as affine functions, i.e.,
\begin{align}
\hat {\cal G}\left( {\bf{ a}} \right) \mathop  \ge \limits_{\left( \rm a \right)}&  \hat {\cal G}\left( {{{\bf{a}}_i}} \right) + 2{\bf{a}}_i^{\rm{T}}({\bf{a}} - {{\bf{a}}_i})
 \buildrel \Delta \over = \hat {\cal G}\left( {{\bf{a}};{{\bf{a}}_i}} \right), \label{eq46}\\
\Upsilon \left( {\bf{z}} \right)   \mathop  \ge \limits_{\left(\rm  b\right)} &  \Upsilon \left( {{{\bf{z}}_i}} \right) + {\nabla _{{z^{\rm{A}}}}}\Upsilon \left( {{\bf z}_i} \right)\left( {{z^{\rm{A}}} - z_{^i}^{\rm{A}}} \right)
  \buildrel \Delta \over =  \Upsilon \left( {{\bf{z}};{{\bf{z}}_i}} \right) \label{eq47},
\end{align}
where $\Upsilon \left( {\bf{z}} \right) $ is defined in \eqref{cons20}, and $
{\nabla _{{z^{\rm{A}}}}}\Upsilon \left( {{{\bf{z}}_i}} \right) =  -  e^{{o_{\max } {- z_{^i}^{\rm{A}}} }}$.
Here,  (a) and (b) are because the Hessian matrix of the convex function is semi-definite. It implies that the second-order remainder term in the Taylor expansion of the function's first-order approximation is non-negative.
Thus, neglecting these non-negative remainder terms right-hand side implies that the right-hand side is smaller than or equal to the left-hand side. Then, the non-convexity of \eqref{cons8} is addressed.
\begin{theorem}\label{theoremD}
 Denoting the solution in the $i$-th iteration of SCA by  ${\bf{V}}_i = \left[ {{\bf{V}}_{1,i}^{\rm{R}},\ldots,{\bf{V}}_{B,i}^{\rm{R}},{\bf{V}}_{1,i}^{\rm{C}},\ldots,{\bf{V}}_{K,i}^{\rm{C}}} \right]$, function ${\cal C}_k \left( {\bf{V}} \right)$ in constraint \eqref{cons8} in the ($i+1$)-th iteration can be approximated by the following  concave lower bound  function:
\begin{align}
{\cal C}_k\left( {\bf{V}} \right) = &\ln \left( {1 + \frac{{{{\cal F}_k}\left( {{{\bf{V}}}} \right)}}{{{{\cal H}_k}\left( {{{\bf{V}}}} \right)}}} \right)\nonumber\\
  \ge &\ln \left( {1 + \frac{{{{\cal F}_k}\left( {{{\bf{V}}_i}} \right)}}{{{{\cal H}_k}\left( {{{\bf{V}}_i}} \right)}}} \right) - \frac{{{{\cal F}_k}\left( {{{\bf{V}}_i}} \right)}}{{{{\cal H}_k}\left( {{{\bf{V}}_i}} \right)}} + 2\frac{{{{\cal X}_k}\left( {{{\bf{V}}_i}} \right){{\cal X}_k}\left( {\bf{V}} \right)}}{{{{\cal H}_k}\left( {{{\bf{V}}_i}} \right)}}\nonumber\\
 &- \frac{{{{\cal F}_k}\left( {{{\bf{V}}_i}} \right)\left( {{{\cal F}_k}\left( {\bf{V}} \right) + {{\cal H}_k}\left( {\bf{V}} \right)} \right)}}{{{{\cal H}_k}\left( {{{\bf{V}}_i}} \right)\left( {{{\cal F}_k}\left( {{{\bf{V}}_i}} \right) + {{\cal H}_k}\left( {{{\bf{V}}_i}} \right)} \right)}}\nonumber\\
  =&\; {\cal C}_k\left( {{\bf{V}};{{\bf{V}}_i}} \right), \label{eq50a}
\end{align}
where ${{{\cal F}_k}\left( {{{\bf{V}}}} \right)}$ and ${{{\cal H}_k}\left( {{{\bf{V}}}} \right)}$ are defined in \eqref{eq33a} and \eqref{eq34a}, respectively, and ${{\cal X}_k}\left( {\bf{V}} \right) = {\left( {{{\cal F}_k}\left( {\bf{V}} \right)} \right)^{\frac{1}{2}}}=\left({\rm{Tr}}\left( {{\bf{G}}_k^{\rm{C}}{\bf{V}}_k^{\rm{C}}} \right)\right)^{\frac{1}{2}}$.
\end{theorem}
\begin{IEEEproof}
Please refer to Appendix \ref{appendixD}.
\end{IEEEproof}

Substituting \eqref{eq46}, \eqref{eq47}, and \eqref{eq50a} into the objective function, constraint \eqref{cons8}, and constraint \eqref{cons20}, respectively, problem \ref{eqP6} in the ($i+1$)-th iteration of SCA can be approximated by
\begin{align}
\mathop {\max }\limits_{{\bf{V}},{\bf{z }},{\bf{a}}}\;&\;{\cal U}\left({\bf{z }}\right)-{\hat \eta }\left({\cal G}\left( {\bf{a}} \right) - \hat {\cal G}\left( {\bf{a}} ;{\bf{a}}_i\right)\right)\tag{\bf P7}\label{eqP7}\\
{\rm{s}}{\rm{.t}}{\rm{.}}\;&z_k^{\rm{C}} \le {\cal C}_k\left( {{\bf{V}};{{\bf{V}}_i}} \right), \forall k, \tag{\rm C26} \label{cons26} \\
&\sum\nolimits_{b = 1}^B {{a_b}{o_b}}  -  \Upsilon \left( {{\bf{z}};{{\bf{z}}_i}} \right) \le 0, \tag{\rm C27} \label{cons27} \\
 &\eqref{cons7},\eqref{cons9}-\eqref{cons14}, \eqref{cons17}-\eqref{cons19}, \eqref{cons25}. \nonumber
\end{align}

\begin{table*}[t]
 \centering
 {
 \begin{tabular}{|c|c||c|c|}
\hline
{Parameters}      &{Value}                                  &{Parameters}            &{Value}\\ \hline
{Speed of light}     &{$c$ = $3\times10^8$ m/s }                      &{Number of subcarriers} &{$M$ = 64}\\ \hline
{Total signal bandwidth} &{$M\Delta_f$ = 10 MHz }&{Subcarrier bandwidth}&{$\Delta_f$ = 156.25 KHz} \\ \hline
 {Elementary OFDM symbol duration} &{$T_o$ = $1/\Delta_f$ = 6.4 us}  &{Cyclic prefix duration}      &{$T_{\rm cp}$ = $\frac{1}{4}T$ = 1.6 us} \\ \hline
  Transmit OFDM symbol duration&{$T$ =  8 us}     &{Number of symbols }          &$L$ = 50\\ \hline
{Communication and sensing weights }    &$w_k^{\rm C}=1/K$, $w_q^{\rm R}=1/Q$& Noise power spectrum density & $N_0=-174$ dBm/Hz \\ \hline
Priority parameters & $\alpha_{\rm C}=0.3$, $\alpha_{\rm R}=0.3$, $\alpha_{\rm A}=0.4$ &Radar cross-section coefficient & ${\zeta _{q}}=1$   \\ \hline
Maximum sensing errors& ${\varepsilon _q^{{\rm D},\max }}=15$, ${\varepsilon _q^{{\rm V},\max }}=1.5$ &  Minimal transmission rate & ${C_k^{\min }}=1.5 $ bits/s/Hz \\ \hline
 \end{tabular} }
\caption{Simulation parameter setup}\label{table1}
\end{table*}

However, this problem is still non-convex due to the rank constraint \eqref{cons10}.
Fortunately, we can apply the same method from Theorem \ref{theoremrank1} to demonstrate that there exists a globally optimal solution with rank at most one for the relaxed formulation of problem \ref{eqP7} when constraint \eqref{cons10} is omitted.
This optimal solution coincides with the optimal solution of problem \ref{eqP7}.
Moreover, due to the convexity of the relaxed problem obtained by ignoring constraint \eqref{cons10}, it can be efficiently solved using the MATLAB toolbox CVX.
Finally, we summarize the details of solving  \ref{eqP6} in Algorithm~\ref{algorithmSCA}

\subsection{Performance Analysis}

\subsubsection{Algorithm Optimality and Convergence}
From \cite{chen2018resource,chen2019intelligent}, we know that the SCA method will eventually converge, as the objective function iteratively increases until it reaches a stationary point. Moreover, since we relax the integer variables and non-convex constraints, it is a suboptimal algorithm.

\subsubsection{Algorithm Complexity} The SCA algorithm is an iterative approach with each iteration solving problem \ref{eqP7}.
Similarly,  from \cite{nesterov1994interior}, we know the complexity of solving  \ref{eqP7} is generally given by ${\cal O}\left(  \widetilde{\cal   N}_C^{1.5} \widetilde {\cal N}_D^{2}\right)$, where $  \widetilde{\cal N}_C=B + 2K + 8Q + \left| {{{\cal S}_o}} \right| + 8$ and $ \widetilde {\cal N}_D=B + K + \left( {B + K{B^2}} \right)N_{{\rm{tx}}}^2 + 2$ represent the number of constraints and the dimension of
optimization variables, respectively.
Therefore, denoting the number of iterations for convergence of the SCA by ${\cal N}_{\rm SCA}$, the corresponding complexity can be generally expressed as  ${\cal O}\left({\cal N}_{\rm SCA} \widetilde{\cal N}_C^{1.5}\widetilde {\cal N}_D^{2}\right)$.

\begin{algorithm}[t]{
\caption{SCA algorithm for  problem \ref{eqP6}}\label{algorithmSCA}
\renewcommand{\baselinestretch}{0.95}
{
Initialize $i=0$,  $ {\bf a}_0$, $ {\bf V}_0$,  $ {\bf z}_0$\;
\Repeat{  the objective function value has converged}{
    Given ${\bf a}_i$, $ {\bf V}_i$, and  ${\bf z}_i$, calculate ${\bf a}_{i+1}$, $ {\bf V}_{i+1}$, and ${\bf z}_{i+1}$
    by solving problem \ref{eqP7}\;
    $i \leftarrow i+1$\;
    }
} }
\end{algorithm}

\begin{figure*}[t]
    \centering
    \begin{subfigure}[b]{0.45\textwidth}
        \centering
        \includegraphics[trim=8 1.7 8 18, clip,width=\textwidth]{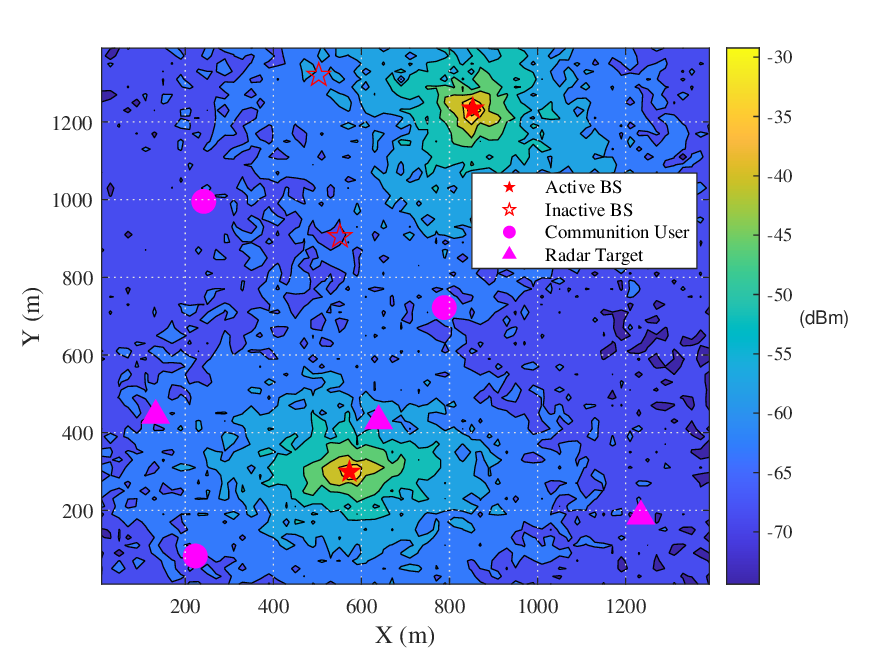}\vspace{-0.2cm}
        \caption{MO-BRB}
        \label{fig:subfig1}
    \end{subfigure}
    \begin{subfigure}[b]{0.45\textwidth}
        \centering
        \includegraphics[trim=8 1.7 8 18, clip,width=\textwidth]{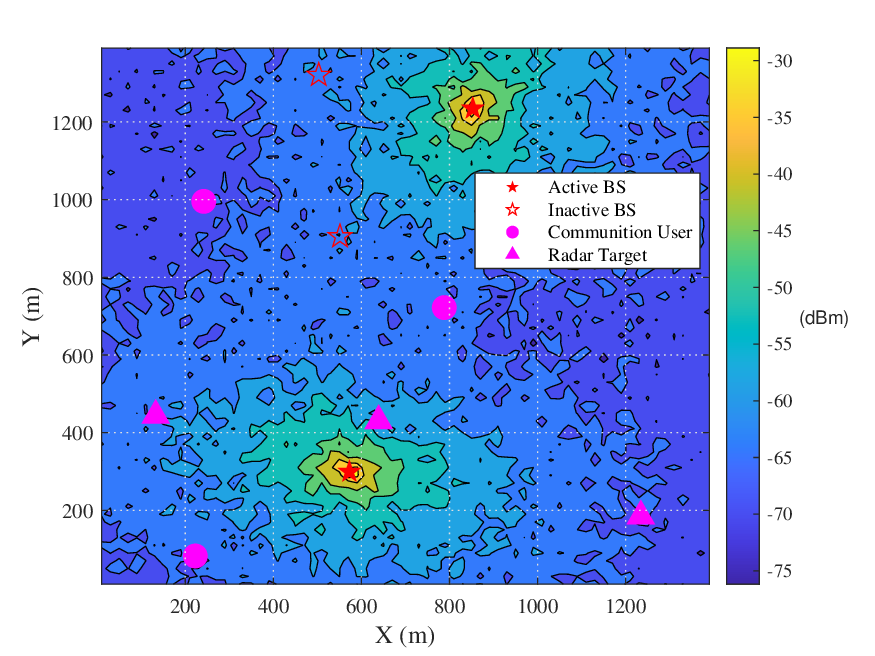}\vspace{-0.2cm}
        \caption{SCA}
        \label{fig:subfig2}
    \end{subfigure}
    \caption{A snapshot of the radiation footprint for the proposed optimal and sub-optimal algorithms: $P=15$ dBm, $B=4$, $Q=3$, $K=3$, $N_{\rm bs}=4$, and $N_{\rm tx}=5$.}
    \label{figSim1}
\end{figure*}

\section{Simulation Results}
In this paper, the parameters for the service provision area and sub-region are defined as follows: $D_x= 1.4$ km, $D_y= 1.4$ km, $\Delta_x= 200$ m, and $\Delta_y= 200$ m,  $X=7$, and $Y=7$, respectively. Then, we consider that the $(x,y)$-th sub-region has the following radiation power constraint:  $I_{xy}^{\max}=I ^{\max}$ for $\left\{ {\left( {x,y} \right)\left| {1 \le x \le \left\lceil {\frac{X}{2}} \right\rceil ,1 \le y \le \left\lceil {\frac{Y}{2}} \right\rceil } \right.} \right\}$ or $\left\{ {\left( {x,y} \right)\left| {\left\lceil {\frac{X}{2}} \right\rceil  \le x \le X,\left\lceil {\frac{Y}{2}} \right\rceil  \le y \le Y} \right.} \right\}$, otherwise  $I_{xy}^{\max}=-10+I ^{\max}$, where $I ^{\max}=-23.01$ dBm.
The carrier frequency is set as 5.89 GHz from IEEE 8022.11p \cite{chen2023impact}. The Rician factor and path propagation loss exponent are set to $\kappa=2$ and $\xi=2.3$, respectively.
The velocity of each target along the x- or y-axis is uniformly distributed between 30 and 50 m/s.
The penalty factor in the SCA method is set to $ {\hat \eta } =15$. We set $p_b^{\rm max}=P$ for $1\le b \le B$.   The BS operation cost is randomly generated from a uniform distribution between $\left[1,4\right]$.
Moreover, the remaining parameters are given in Table~\ref{table1}~\cite{nguyen2017delay}.
The presented results are the average of 200 Monte Carlo simulations.
We also compare the proposed algorithms with the following baselines.
In the  ``Maximum'' baseline, we randomly select the maximum $N_{\rm{bs}}$ BSs with equal probability from the entire set and activate all of them, and then optimize the remaining beamforming coordination using SCA.
In the  ``Random'' baseline, we randomly activate each BS with a probability of 0.5. If the number of activated BSs exceeds the maximum $N_{\rm bs}$, we randomly select $N_{\rm{bs}}$ BSs with equal probability from the initially activated ones, again employing  SCA for beamforming optimization.
Inspired by \cite{zhang2018joint}, we define the BS activation priorities as ${\dot {\cal P}_b} = \frac{1}{{{o_b}}}\left( {\sum\nolimits_{k = 1}^K {{{\left\| {{\bf{g}}_{bk}^{\rm{C}}} \right\|}^2} + } \sum\nolimits_{q = 1}^Q {{{\left\| {{\bf g}_{bq}^{\rm{R}}} \right\|}^2}} } \right)$.
Then, in the ``Priority-Maximum" baseline, we activate $N_{\rm{bs}}$ BSs with the highest priorities and subsequently optimize the beamforming coordination using SCA.
In the ``Priority-Half" baseline, we activate $\left\lfloor {\frac{{{N_{{\rm{bs}}}}}}{2}} \right\rfloor$ BSs with the highest priorities, and then optimize the beamforming coordination using SCA.
Furthermore, since the optimal MO-BRB requires exponential computational complexity, we only show the performance of this method with small dimensions to validate the effectiveness of the proposed suboptimal algorithms.

Fig. \ref{figSim1} shows the radiation footprint of all BSs within the
studied service provisioning area.
The figure indicates that, by strategically activating a subset of BS and optimizing
the beamforming, the radiation footprint can be effectively reduced below the threshold,  validating  the effectiveness
of the proposed algorithms.
Additionally, areas with high device density exhibit higher radiation levels compared to other regions, owing to directional signals providing S\&C services.

\begin{figure}[t]
\center
\includegraphics[trim=15 1.7 15 18, clip,width=0.45\textwidth]{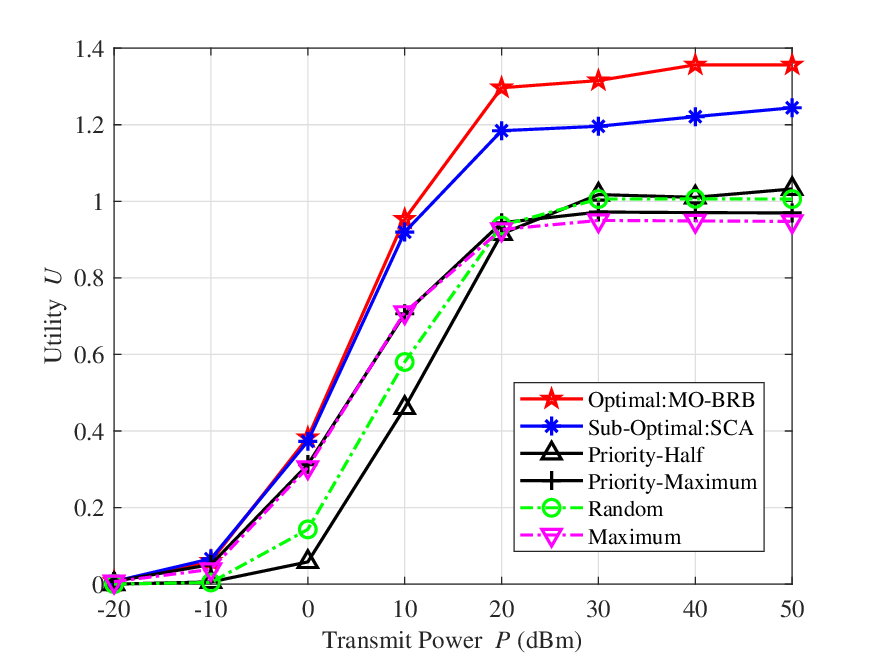}
\caption{The system utility versus transmission power:  $B=4$, $Q=2$, $K=2$, $N_{\rm bs}=3$, and $N_{\rm tx}=4$.} \label{figSim3}
\end{figure}

\begin{figure}[t]
\center
\includegraphics[trim=15 1.7 15 18, clip,width=0.45\textwidth]{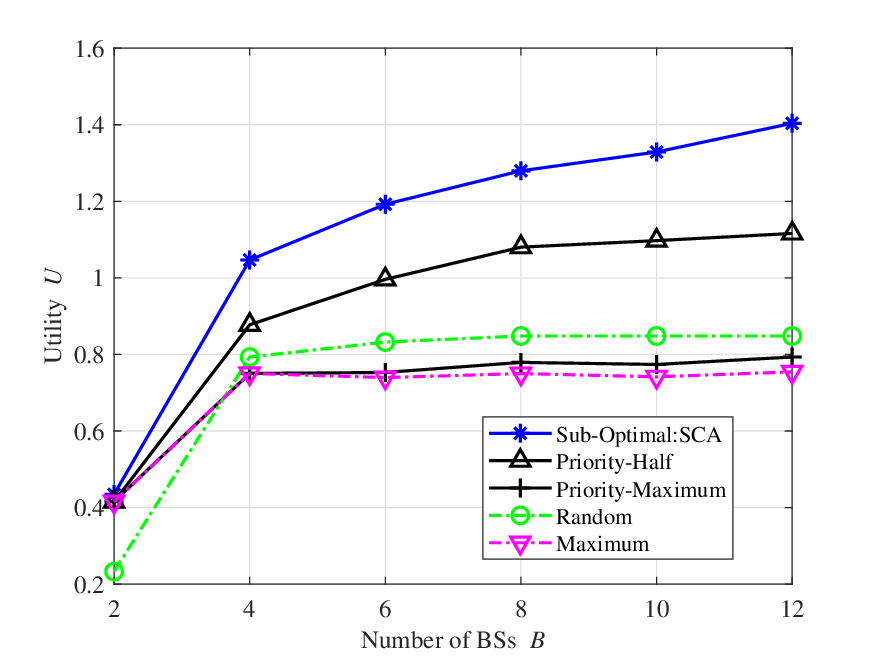}
\caption{The system utility versus the number of BSs:  $P = 15$ dBm,  $Q=2$, $K=2$, $N_{\rm bs}=5$, and $N_{\rm tx}=4$.} \label{figSim4}
\end{figure}

\begin{figure}[t]
\center
\includegraphics[trim=15 1.7 15 18, clip,width=0.44\textwidth]{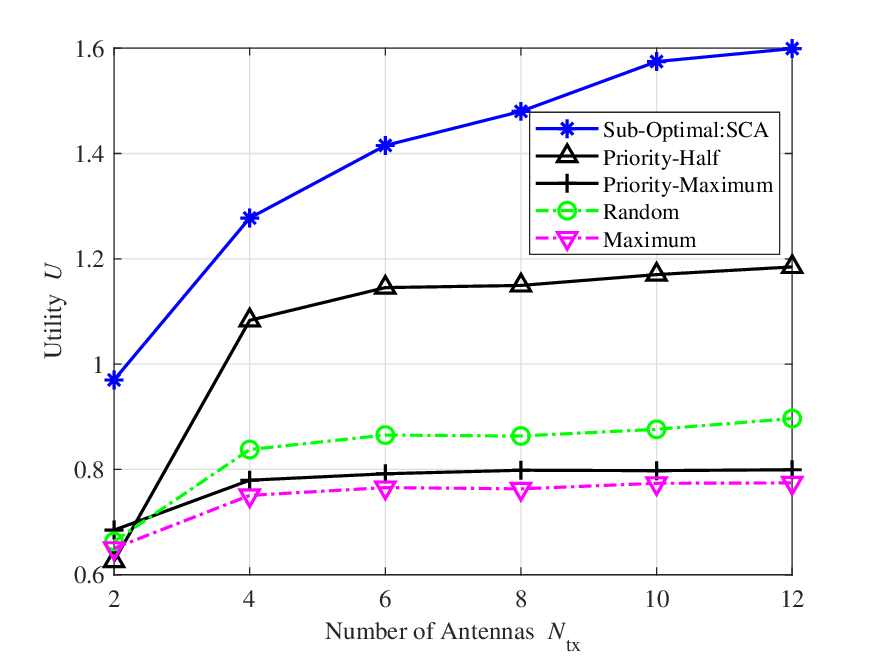}
\caption{The system utility versus the number of transmit antennas: $P = 15$ dBm, $B=8$, $Q=2$, $K=2$, and $N_{\rm bs}=5$.} \label{figSim5}
\end{figure}

Fig.~\ref{figSim3} shows the impact of the maximum transmission power of each BS on system utility.
Initially, all algorithms yield zero utility due to insufficient power to meet the minimum rate and sensing accuracy requirements, making the problem infeasible. As power increases, the improved signal-to-noise ratio enhances both metrics, resulting in a rise in utility.
Eventually, the utility saturates because additional power cannot be effectively utilized due to the radiation constraint.
The proposed sub-optimal algorithm closely approaches the optimal performance and outperforms other benchmarks, demonstrating its effectiveness.

\begin{figure}[t]
    \centering
    \begin{subfigure}[b]{0.45\textwidth}
        \centering
       \includegraphics[trim=15 1.7 15 18, clip,width=\textwidth]{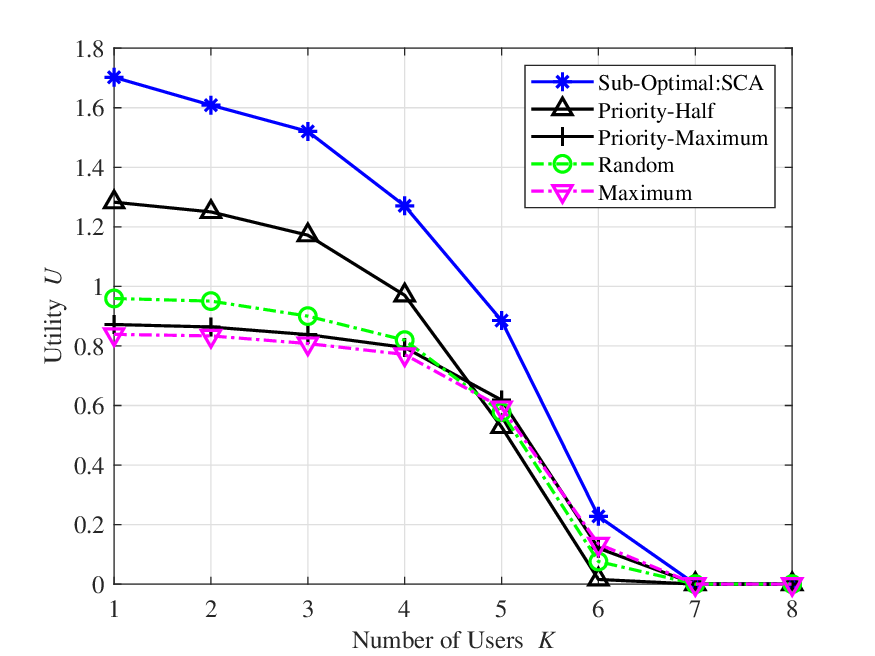}
        \caption{Utility}
        \label{figSim61}
    \end{subfigure}
    \begin{subfigure}[b]{0.45\textwidth}
        \centering
\includegraphics[trim=15 1.7 15 18, clip,width=\textwidth]{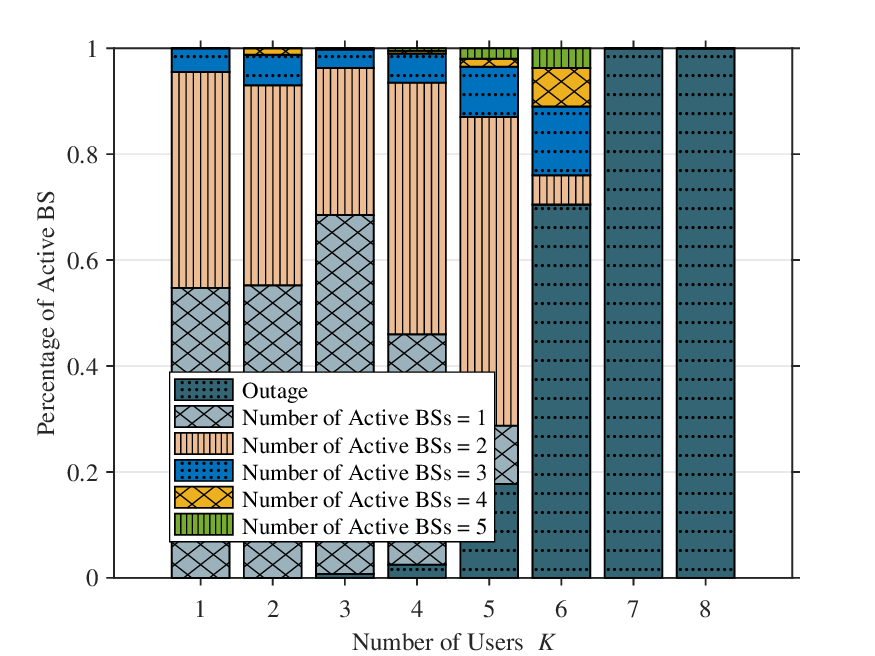}
        \caption{Occurrence Percentage}
        \label{figSim62}
    \end{subfigure}
    \caption{The system performances versus the number of users/targets: $P = 15$ dBm, $B=8$, $Q=K$, $N_{\rm bs}=5$, and $N_{\rm tx}=8$.} \label{figSim6}
    \end{figure}

\begin{figure}[t]
\center
\includegraphics[trim=15 1.7 15 18, clip, width=0.45\textwidth]{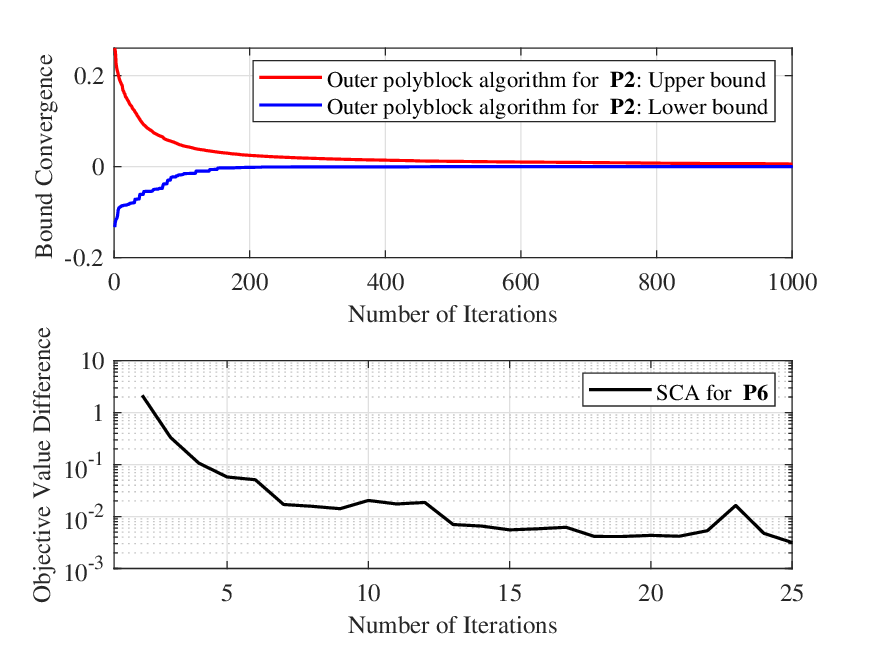}
\caption{Convergence performances of outer polyblock and SCA algorithms: $P = 20$ dBm, $B=3$, $Q=2$, $K=2$, $N_{\rm bs}=3$, and $N_{\rm tx}=4$.} \label{figSim9} 
\end{figure}

Fig. \ref{figSim4}  illustrates the impact of the number of BSs on the system performance.
It can be observed that the utility functions for all algorithms increase as the number of BSs increases.
This is because the increased number of BSs enhances the BS activation flexibility, thereby offering greater freedom of optimization.
Additionally, it boosts array gain and enables effective beamforming coordination across BSs to enhance the system's utility.
Moreover, the performance gap between the proposed algorithm and the benchmark algorithms widens with the increase in the number of BSs,  validating the importance of BS activation and beamforming coordination.

Fig. \ref{figSim5} illustrates the impact of the number of antennas on the system utility.
It is observed that the system utility for all algorithms improves as the number of antennas increases. This is because a larger number of antennas provides greater spatial diversity and higher array gains, thereby enabling more effective beamforming coordination among multiple BSs. Furthermore, the performance gap between the proposed algorithm and the benchmark algorithms widens with the increase in the number of antennas, further highlighting the importance of joint BS activation and beamforming coordination.

Fig. \ref{figSim6} shows the impact of the number of users on the system performance.
As the number of users increases, the system utilities of all algorithms decline, eventually dropping to zero.
This is because more resources are demanded to ensure coverage for distant users, thereby reducing system utility.
Comparing baselines based on ``Random'' and ``Maximum'', with fewer users, activating a smaller number of BSs results in higher system utility compared to activating all stations. This is because the current BSs can meet the demands of fewer users without requiring additional BSs to increase costs.
Furthermore, as the number of users increases, it requires a higher number of activated BSs to ensure S\&C service demands.
However, due to the constraints on maximum power and the number of active BSs, it becomes increasingly challenging to meet these requirements.
Consequently, the optimization may become infeasible, thus increasing the outage probability.

Fig. \ref{figSim9} shows the convergence performances of the proposed algorithms. For the outer polyblock algorithm in the MO, the difference between the upper/lower bounds and the converged value is evaluated at each iteration. For the SCA algorithm, the absolute difference of the objective function between two successive iterations is measured. The results show that the MO algorithm converges after hundreds of iterations, while the SCA algorithm achieves convergence within a threshold of $10^{-2}$ in just a few iterations, demonstrating its low complexity.

\section{Conclusion}

This paper proposes an interference-suppressed and cost-optimized cell-free cooperative distributed ISAC network by opportunistically leveraging distributed resources to address individual S\&C demands.
Specifically, we conceive a radiation footprint control mechanism to autonomously reduce interference across the entire service area and protect other networks sharing the spectrum.
Then, we develop a joint BS activation and beamforming coordination scheme to dynamically orchestrate spatial beams from appropriately activated BSs,  thus achieving high-quality service provisioning and cost reduction.
Next, we formulate a cost-efficient utility maximization problem while considering individual S\&C service demands and location-specific radiation footprint constraints.
Then, we develop an efficient  MO-BRB algorithm to find its optimal solution and a low-complexity SCA method to find its near-optimal solution.
Simulations reveal that the coordinated scheme---by integrating BS activation and beamforming---significantly outperforms the non-coordinated approaches.
This advantage is particularly pronounced in scenarios featuring dense BS deployment, extensive antenna arrays, and higher transmission power levels.
Finally, while this work relies on partial statistical channel knowledge within the radiation footprint control regions and uses a proportional fairness metric to evaluate heterogeneous system performance, future research may explore achieving effective control with reduced channel knowledge and develop more refined or comprehensive metrics to better balance the diverse requirements of sensing, communication, and system cost.

\appendix

\subsection{Proof of Theorem \ref{theorem1}}\label{appendix1}

In this part, we derive the FIM  in \eqref{eq17}. To begin with, the first derivative of $\log {P}\left( {{{{\bf{\bar r}}}_q}\left| {{{\bm{\omega }}_q}} \right.} \right)$ with respect to ${\tau _{bb'q}}$ is
\begin{align}
&\frac{{\partial \log P\left( {{{{\bf{\bar r}}}_q}\left| {{{\bm{\omega }}_q}} \right.} \right)}}{{\partial {\tau _{bb'q}}}} = \frac{1}{{{\sigma ^2}}} \sum\limits_{m = 0}^{M - 1} {\sum\limits_{l = 0}^{L - 1} {\left( {\left( {\bar r_{b'q}^{ml} - \sum\limits_{b = 1}^B {\mu _{bb'q}^{ml}} } \right)} \right.} } \nonumber\\
 &\times \left. {{\frac{{\partial {{\left( {\mu _{bb'q}^{ml}} \right)}^*}}}{{\partial {\tau _{bb'q}}}}} + {{\left( {\bar r_{bq}^{ml} - \sum\nolimits_{b = 1}^B {\mu _{bb'q}^{ml}} } \right)}^*}\frac{{\partial \mu _{bb'q}^{ml}}}{{\partial {\tau _{bb'q}}}}} \right).
 \end{align}
Then, the ($i,j$)-th element ${\left[ {{{\bf{J}}_q^{\tau \tau }}} \right]_{ij}}$ can be rewritten as:
 \begin{align}
&{\mathbb{E}}\left( { - \frac{{{\partial ^2}\log P\left( {{{\bf{\bar r}}_q}\left| {{{\bm{\omega }}_q}} \right.} \right)}}{{\partial {\tau _{bb'q}}\partial {\tau _{\tilde b\tilde b'q}}}}} \right)\nonumber\\
 = &\left\{ {\begin{array}{*{20}{l}}
{\frac{{2{\mathbb{E}}    \left( {\sum\limits_{m,l} {{\rm{Re}}\left\{ {\frac{{\partial \mu _{bb'q}^{ml}}}{{\partial {\tau _{bb'q}}}}\frac{{\partial {{\left( {\mu _{\tilde bb'q}^{ml}} \right)}^*}}}{{\partial {\tau _{\tilde bb'q}}}}} \right\}} } \right)}}{{{\sigma ^2}}},\tilde b' = b',}\\
{0,\tilde b' \ne b',}
\end{array}} \right.
\nonumber \\
 = &\left\{ {\begin{array}{*{20}{l}}
{\frac{{2{\mathbb{E}}\left( {\sum\limits_{m,l} {{\rm{Re}}\left\{ {4{\pi ^2}{m^2}\Delta _f^2\bar \xi _{b'q}^2\bar s_{bq}^{ml}{{\left( {\bar s_{\tilde bq}^{ml}} \right)}^*}{e^{j2\pi \Delta _{ij}^{ml}}}} \right\}} } \right)}}{{{\sigma ^2}}},\tilde b' = b',}\\
{0,\tilde b' \ne b',}
\end{array}} \right.\nonumber\\
  \approx &\left\{ {\begin{array}{*{20}{l}}
{\frac{{{\mathbb{E}}\left( {\sum\limits_{m,l} {8{\pi ^2}{m^2}\Delta _f^2\bar \xi _{b'q}^2{{\left| {\bar s_{bq}^{ml}} \right|}^2}} } \right)}}{{{\sigma ^2}}},i = j,}\\
{0,i \ne j,}
\end{array}} \right.\nonumber\\
 =& \left\{ {\begin{array}{*{20}{l}}
{\beta _{b'q}^{\tau \tau }{\lambda _{bq}},i = j,}\\
{0,i \ne j,}
\end{array}} \right. \label{eq51}
  \end{align}
where   $\Delta _{ij}^{ml} =  {\left( {{f_{bb'q}} - {f_{b\tilde b'q}}} \right)lT - m{\Delta _f}\left( {{\tau _{bb'q'}} - {\tau _{b\tilde b'q}}} \right)} $ and $\beta _{bq}^{\tau \tau }$ is given in \eqref{eqp17}. Similarly, we have
 \begin{align}
{\mathbb{E}}\left( {\frac{{{\partial ^2}\log P\left( {{{\bf{\bar r}}_q}\left| {{{\bf{\omega }}_q}} \right.} \right)}}{{\partial {f_{bb'q}}\partial {f_{\tilde b\tilde b'q}}}}} \right) \approx& \left\{ {\begin{array}{*{20}{l}}
{\frac{{{\mathbb{E}}\left( {\sum\limits_{m,l} {8{\pi ^2}{l^2}{T^2}\bar \xi _{b'q}^2{{\left| {\bar s_{bq}^{ml}} \right|}^2}} } \right)}}{{{\sigma ^2}}},i = j}\\
{0,i \ne j}
\end{array}} \right.\nonumber\\
 \approx& \left\{ {\begin{array}{*{20}{l}}
{\beta _{b'q}^{ff}{\lambda _{bq}},i = j}\\
{0,i \ne j}
\end{array},} \right.\label{eq52}\\\
{\mathbb{E}}\left( {\frac{{{\partial ^2}\log P\left( {{{\bf{\bar r}}_q}\left| {{{\bf{\omega }}_q}} \right.} \right)}}{{\partial {\tau _{bb'q}}\partial {f_{\tilde b\tilde b'q}}}}} \right)
 \approx& \left\{ {\begin{array}{*{20}{l}}
{-\beta _{b'q}^{\tau f}{\lambda _{bq}},i = j}\\
{0,i \ne j}
\end{array},} \right.\label{eq53}
\end{align}
where $\beta _{bq}^{f f }$ and  $\beta _{bq}^{\tau f }$ are defined in \eqref{eqp18} and \eqref{eqp19}, respectively.

Then, we ignore the coupling between distance and velocity, simplifying the non-diagonal sub-matrix of ${\bf J}_q$ to a zero matrix.
Upon substituting  \eqref{eq51}, \eqref{eq52}, and \eqref{eq53} into   \eqref{eq11} and performing subsequent  algebraic operations, we have \eqref{eq17}.

\subsection{Prof of Theorem \ref{theoremrank1}}\label{appendix2}

Let $\widehat{\bf{V}} = \left[ {\hat{\bf{V}}_{1,}^{\rm{R}},\ldots,\hat{\bf{V}}_B^{\rm{R}},\hat{\bf{V}}_1^{\rm{C}},\ldots,\hat{\bf{V}}_K^{\rm{C}}} \right]$ be the globally optimal solution to problem \ref{eqp5}. Then, we can construct a new solution ${\bf{V}}^\star = \left[ {{\bf{V}}_{1}^{{\rm R}^\star},\ldots,{\bf{V}}_B^{{\rm R}^\star},{\bf{V}}_1^{{\rm C}^\star},\ldots,{\bf{V}}_K^{{\rm C}^\star}} \right]$, i.e.,
 \begin{align}
{\bf{w}}_k^{{{\rm{C}}^ \star }} &= {\left( {{{\left( {{\bf{g}}_k^{\rm{C}}} \right)}^{\rm H}}{\bf{\hat V}}_k^{\rm{C}}{\bf{g}}_k^{\rm{C}}} \right)^{ - \frac{1}{2}}}{\bf{\hat V}}_k^{\rm{C}}{\bf{g}}_k^{\rm{C}}, \label{eqa57}
{\bf{V}}_k^{{{\rm{C}}^ \star }} &= {\bf{w}}_k^{{{\rm{C}}^ \star }}{\left( {{\bf{w}}_k^{{{\rm{C}}^ \star }}} \right)^{\rm H}},\\
{\bf{V}}_b^{{{\rm{R}}^ \star }} &= {\bf{\hat V}}_b^{\rm{R}} + \Delta {\bf{V}}_b^{\rm{R}}, \label{eqa58}
\end{align}
where $\Delta {\bf{V}}_b^{\rm{R}}$ is a submatrix composed of the rows and columns from $(b-1)N_{\rm tx}+1$ to $bN_{\rm tx}$ of matrix   ${\sum\nolimits_{k = 1}^K {\left( {{\bf{\hat V}}_k^{\rm{C}} - {\bf{V}}_k^{{{\rm{C}}^ \star }}} \right)} }$. Obviously, the new solution ${\bf{V}}^\star $ satisfies     constraint \eqref{cons10}.

From \eqref{eqa57}, for arbitrary vector ${\bf h}\in{\mathbb C}^{BN_{\rm tx}\times1}$, we have
\begin{align}
{{\bf{h}}^{\rm H}}\left( {{\bf{\hat V}}_k^{\rm{C}} - {\bf{V}}_k^{{{\rm{C}}^ \star }}} \right){\bf{h}} = {{\bf{h}}^{\rm H}}{\bf{\hat V}}_k^{\rm{C}}{\bf{h}} - \frac{{{{\left| {{{\bf{h}}^{\rm H}}{\bf{\hat V}}_k^{\rm{C}}{\bf{g}}_k^{\rm{C}}} \right|}^2}}}{{{{\left( {{\bf{g}}_k^{\rm{C}}} \right)}^{\rm H}}{\bf{\hat V}}_k^{\rm{C}}{\bf{g}}_k^{\rm{C}}}}.
\end{align}
Based on the Cauchy-Schwarz inequality, we have
\begin{align}
\left[ {{{\bf{h}}^{\rm H}}{\bf{\hat V}}_k^{\rm{C}}{\bf{h}}} \right]\left[ {{{\left( {{\bf{g}}_k^{\rm{C}}} \right)}^{\rm H}}{\bf{\hat V}}_k^{\rm{C}}{\bf{g}}_k^{\rm{C}}} \right] \ge {\left| {{{\bf{h}}^{\rm H}}{\bf{\hat V}}_k^{\rm{C}}{\bf{g}}_k^{\rm{C}}} \right|^2}.
\end{align}
Consequently, it holds that
\begin{align}
{{\bf{h}}^{\rm H}}\left( {{\bf{\hat V}}_k^{\rm{C}} - {\bf{V}}_k^{{{\rm{C}}^ \star }}} \right){\bf{h}}  \ge  0,{\bf{\hat V}}_k^{\rm{C}} \succeq  {\bf{V}}_k^{{{\rm{C}}^ \star }},\forall k,
\end{align}
and
\begin{align}
{{\cal H}_k}\left( {{{\bf{V}}^ \star }} \right)& = \sum\nolimits_{k' \ne k}^K {{{\left( {{\bf{g}}_k^{\rm{C}}} \right)}^{\rm H}}{\bf{V}}_{k'}^{{{\rm{C}}^ \star }}{\bf{g}}_k^{\rm{C}}} \nonumber\\
 & \le  \sum\nolimits_{k' \ne k}^K {{{\left( {{\bf{g}}_k^{\rm{C}}} \right)}^{\rm H}}{\bf{\hat V}}_{k'}^{\rm{C}}{\bf{g}}_k^{\rm{C}}}  = {{\cal H}_k}\left( {{\bf{\hat V}}}  \right), \\
{{\cal F}_k}\left( {{{\bf{V}}^ \star }} \right)& = {\left( {{\bf{g}}_k^{\rm{C}}} \right)^{\rm H}}{\bf{V}}_{k'}^{{{\rm{C}}^ \star }}{\bf{g}}_k^{\rm{C}} = {\left( {{\bf{g}}_k^{\rm{C}}} \right)^{\rm H}}{\bf{w}}_k^{{{\rm{C}}^ \star }}{\left( {{\bf{w}}_k^{{{\rm{C}}^ \star }}} \right)^{\rm H}}{\bf{g}}_k^{\rm{C}}\nonumber\\
 &= {\left( {{\bf{g}}_k^{\rm{C}}} \right)^{\rm H}}{\bf{\hat V}}_k^{\rm{C}}{\bf{g}}_k^{\rm{C}}= {{\cal F}_k}\left( {\widehat {\bf{V}}} \right).
\end{align}

Then,   we know
\begin{align}
\frac{{{{\cal F}_k}\left( {\widehat {\bf{V}}} \right)}}{{{{\cal H}_k}\left( {\widehat {\bf{V}}} \right) + {\sigma _k}}} \le \frac{{{{\cal F}_k}\left( {{{\bf{V}}^ \star }} \right)}}{{{{\cal H}_k}\left( {{{\bf{V}}^ \star }} \right) + {\sigma _k}}}. \label{eqapp1}
\end{align}
Similarly,  we can demonstrate the following equations: in constraint \eqref{cons11}, we have ${\cal P}_b\left({{\widehat {\bf{V}}}}\right)={\cal P}_b\left({{{\bf{V}}^ \star }}\right)$. Moreover, constraints \eqref{cons17} and \eqref{cons18} are also satisfied by ${\bf{V}}^\star$.
It follows that the solution ${\bf{V}}^\star$ satisfying the rank constraint \eqref{cons10} can maintain the system performance achieved by the globally optimal solution of $\widehat {\bf{V}}$. Finally,   Theorem \ref{theoremrank1} is proved.
\subsection{Proof of Theorem \ref{theoremD}}\label{appendixD}

Firstly,   we show that the function $f(x,y) = -\ln(1 - |x|^2/y)$ is convex for  $0 < y < |x|^2$.
Specifically,
since $-\ln(1 - z)$ increases and is convex  of $z$, and $z = |x|^2/y$ is a convex function of $(x,y)$ for $0 < y < |x|^2$, it follows that $f(x,y)$ is convex \cite{chen2019intelligent}.
 According to the first-order Taylor series expansion of $f\left( {x,y} \right) $ at $({\tilde x},{\tilde y})$, we have
\begin{align}
&f\left( {x,y} \right) \ge f\left( {\tilde x,\tilde y} \right) + {\nabla _{\tilde x}}f\left( {x,\tilde y} \right)\left( {x - \tilde x} \right) + {\nabla _{\tilde y}}f\left( {\tilde x,y} \right)\left( {y - \tilde y} \right)\nonumber\\
& = f\left( {\tilde x,\tilde y} \right) + 2\frac{{\Re \left\{ {\tilde x\left( {x - \tilde x} \right)} \right\}}}{{\tilde y - {\left| {\tilde x} \right|}^2}} - \frac{{{{\left| {\tilde x} \right|}^2}}}{{\tilde y\left( {\tilde y - {{\left| {\tilde x} \right|}^2}} \right)}}\left( {y - \tilde y} \right), \label{eqa65}
\end{align}
where ${\nabla _{\tilde x}}f\left( {x,\tilde y} \right)$ and ${\nabla _{\tilde y}}f\left( {\tilde x,y} \right)$ are the constant gradients of $f\left( {x,y} \right) $ at $({\tilde x},{\tilde y})$ with respect to $x$ and $y$, respectively.

Next, we have $\ln (1 + \frac{{{{\left| x \right|}^2}}}{c}) =  - \ln (1 - \frac{{{{\left| x \right|}^2}}}{y})$ if $y=c+\left| x \right|^2$.
With \eqref{eqa65} and denoting ${\tilde c}={\tilde y}-{ \left| {\tilde x} \right|}^2$, it follows that
\begin{align}
\ln (1 + \frac{{{{\left| x \right|}^2}}}{c}) & \ge \ln ( {1 + \frac{{{{\left| {\tilde x} \right|}^2}}}{{\tilde c}}} ) + 2\frac{{\Re \left\{ {\tilde xx} \right\}}}{{\tilde c}} \nonumber\\
&- \frac{{{{\left| {\tilde x} \right|}^2}}}{{\tilde c( {\tilde c + {{\left| {\tilde x} \right|}^2}} )}}( {c + {{\left| { x} \right|}^2}} ) - \frac{{{{\left| {\tilde x} \right|}^2}}}{{\tilde c}}.
\end{align}

Finally, letting
 \begin{align}
x &= {{\cal X}_k}\left( {\bf{V}} \right) = {\left( {{{\cal F}_k}\left( {\bf{V}} \right)} \right)^{\frac{1}{2}}},c = {{\cal H}_k}\left( {\bf{V}} \right),\\
\tilde x &= {{\cal X}_k}\left( {{{\bf{V}}_i}} \right) = {\left( {{{\cal F}_k}\left( {{{\bf{V}}_i}} \right)} \right)^{\frac{1}{2}}},\tilde c = {{\cal H}_k}\left( {{{\bf{V}}_i}} \right),
 \end{align}
we have \eqref{eq50a}. This completes the proof of Theorem \ref{theoremD}.

\ifCLASSOPTIONcaptionsoff

\fi
  \bibliography{SPT}
\bibliographystyle{IEEEtran}

\end{document}